\documentclass[
aps,
prb,
reprint,
showkeys,
twocolumn,
superscriptaddress,
]{revtex4-2}

\usepackage{natbib}
\usepackage{color}
\usepackage{textcomp}
\usepackage{amsmath}
\usepackage[charter]{mathdesign}
\usepackage{easyReview}
\usepackage{microtype} 
\usepackage{upgreek}
\newcommand{\ped}[1]{\ensuremath{_{\rm #1}}}
\newcommand{\apex}[1]{\ensuremath{^{\rm #1}}}
\newcommand{\Tone}{$T_{1}^{\,-1}$}

\newcommand{\musr}{$\upmu$SR}
\definecolor{link}{RGB}{57,106,177}
\definecolor{blue}{RGB}{0,0,0}
\definecolor{darkgreen}{RGB}{0,128,0}
\usepackage{float}
\usepackage[caption = false]{subfig}
\usepackage{graphicx}
\usepackage{hyperref}
\hypersetup{
	colorlinks=true,
	linkcolor=link,
	citecolor=link,
	urlcolor=link
}

\newcommand{\tcb}[1]{\textcolor{blue}{#1}}

\begin{document}

\title{Superconductivity induced by gate-driven hydrogen intercalation
\texorpdfstring{\\}{} 
in the charge-density-wave compound 
\texorpdfstring{1$T$-TiSe$_2$}{1T-TiSe2}}
	
\author{Erik Piatti}
\email{erik.piatti@polito.it}
\affiliation{Department of Applied Science and Technology, Politecnico di Torino, I-10129 Torino, Italy}
\author{Giacomo Prando}
\affiliation{Dipartimento di Fisica, Universit\`a degli Studi di Pavia, I-27100 Pavia, Italy}
\author{Martina Meinero}
\affiliation{Consiglio Nazionale delle Ricerche-SPIN, I-16152 Genova, Italy}
\affiliation{Department of Physics, Università di Genova, I-16146 Genova, Italy}
\author{Cesare Tresca}
\affiliation{Department of Physical and Chemical Sciences, Università degli Studi dell’Aquila, I-67100 L’Aquila, Italy}
\affiliation{SPIN-CNR, Università degli Studi dell'Aquila, I-67100 L'Aquila, Italy}
\author{Marina Putti}
\affiliation{Consiglio Nazionale delle Ricerche-SPIN, I-16152 Genova, Italy}
\affiliation{Department of Physics, Università di Genova, I-16146 Genova, Italy}
\author{Stefano Roddaro}
\affiliation{Istituto Nanoscienze-CNR, NEST and Scuola Normale Superiore, I-56127 Pisa, Italy}
\author{Gianrico Lamura}
\affiliation{Consiglio Nazionale delle Ricerche-SPIN, I-16152 Genova, Italy}
\author{Toni Shiroka}
\affiliation{Laboratory for Muon-Spin Spectroscopy, Paul Scherrer Institut, CH-5232 Villigen PSI, Switzerland}
\affiliation{Laboratorium f{\"u}r Festk{\"o}rperphysik, ETH Z{\"u}rich, CH-8093 Zurich, Switzerland}
\author{Pietro Carretta}
\affiliation{Dipartimento di Fisica, Universit\`a degli Studi di Pavia, I-27100 Pavia, Italy}
\author{Gianni Profeta}
\affiliation{Department of Physical and Chemical Sciences, Università degli Studi dell’Aquila, I-67100 L’Aquila, Italy}
\affiliation{SPIN-CNR, Università degli Studi dell'Aquila, I-67100 L'Aquila, Italy}
\author{Dario Daghero}
\affiliation{Department of Applied Science and Technology, Politecnico di Torino, I-10129 Torino, Italy}
\author{Renato S. Gonnelli}
\email{renato.gonnelli@polito.it}
\affiliation{Department of Applied Science and Technology, Politecnico di Torino, I-10129 Torino, Italy}

\begin{abstract}
Hydrogen (H) plays a key role in the near-to-room temperature superconductivity of hydrides at megabar pressures.
This suggests that H doping could have similar effects on the electronic and phononic spectra of materials at ambient pressure as well.
Here, we demonstrate the non-volatile control of the electronic ground state of titanium diselenide ($1T$-TiSe$_2$) via ionic liquid gating-driven H intercalation. This protonation induces a superconducting phase, observed together with a charge-density wave through most of the phase diagram, with nearly doping-independent transition temperatures.
The H-induced superconducting phase is possibly gapless-like and multi-band in nature, in contrast with those induced in TiSe$_2$ via copper, lithium, and electrostatic doping.
This unique behavior is supported by \emph{ab initio} calculations showing that high concentrations of H dopants induce a full reconstruction of the bandstructure, although with little coupling between electrons and high-frequency H phonons.
Our findings provide a promising approach for engineering the ground state of transition metal dichalcogenides and other layered materials via gate-controlled protonation.

\bigskip
\textbf{Cite this article as:} 

Piatti, E.; Prando, G.; Meinero, M. \textit{et al.} Superconductivity induced by gate-driven hydrogen intercalation in the charge-density-wave compound 1$T$-TiSe$_2$. \textit{Commun. Phys.} \textbf{6}, 202 (2023).
%
%

\textbf{DOI:} \href{https://doi.org/10.1038/s42005-023-01330-w}{10.1038/s42005-023-01330-w}
\end{abstract}

\maketitle
	
\section{Introduction}

The recent observation of near-to-room-temperature superconductivity in hydrides under high pressure\,\cite{DrozdovNature2015, DrozdovNature2019} has demonstrated that a superconducting (SC) state with high critical temperature does not necessarily require an unconventional electron-electron coupling mechanism. The key requirements here are high phonon frequencies, such as those associated to the hydrogen (H) vibration modes, a strong coupling of these modes to the electronic states at the Fermi level, and a non-conventional structural environment (in the specific case, stabilized by pressure) that favours such a coupling. A still-open question is whether these same ingredients can be used to design new SC materials at ambient pressure -- with properties of interest for applications even though at lower temperatures -- or a different paradigm must be considered when designing H-rich superconductors at low pressures. This point is especially relevant in light of the recently-claimed attainment of room-temperature superconductivity in lutetium hydride at relatively low pressures\,\cite{Dias2023}.

One of the routes to create H-rich materials is ionic liquid gating-induced protonation. As this technique makes it possible to insert H atoms in crystallographic positions unattainable by a conventional synthesis approach, it has found notable applications in oxides\,\cite{LuNature2017, JoAFM2018, RafiqueNanoLett2019, LiNatCommun2020, WangAFM2021, ShenPRX2021} and demonstrated a great potential in layered materials as well\,\cite{CuiCPL2019, MengPRB2022}.
Among the latter, transition metal dichalcogenides (TMDs) are particularly interesting because, despite their structural simplicity, they exhibit a variety of quantum phases\,\cite{ManzeliNatRevMater2017, ChoiMT2017} -- charge-density wave (CDW), ferromagnetism, Mott insulating state, etc. -- that may be tuned by protonation. In some cases, TMDs display a topological order, hence, once made superconducting, they may host Majorana fermions\,\cite{FuPRL2008, LiNatCommun2021}. TMDs are also easy to handle and can be used to fabricate electronic devices\,\cite{WangNatNano2012, FioriNatNano2014, ZhuNanoscale2017}, even printed ones\,\cite{LiNatMater2021, PiattiNatElectron2021}. Hence, by tailoring their physical properties, one can realize good and affordable prototypes for various quantum-technology applications, superconducting electronics, or quantum computing\,\cite{LianPNAS2018}. 
In this context, the archetypal TMD compound titanium diselenide (1$T$-TiSe\ped{2}) is an ideal candidate to benchmark the ionic-gating-induced protonation technique, since it exhibits strong electron-electron correlations and a variety of intriguing quantum phases\,\cite{WangNatNano2012}, akin to those found in other layered materials such as iron-based compounds\,\cite{StewartRMP2011, FernandesNature2022}, cuprates\,\cite{ShenMT2008}, and heavy-fermion systems\,\cite{StewartRMP1984}. Furthermore, the electronic ground state of 1$T$-TiSe\ped{2} is readily tunable by a variety of different methods, including chemical intercalation\,\cite{MorosanNatPhys2006, MorosanPRB2010, SatoJPSJ2017}, applied pressure\,\cite{KusmartsevaPRL2009}, and electrostatic gating\,\cite{LiNature2016, LiNanoLett2019}.

In this work, we show that protonation of TMDs induced by ionic liquid gating is a powerful and unique doping technique to control their electronic ground state. In particular, we demonstrate that it allows a robust and non-volatile control over CDW and SC in $1T$-TiSe\ped{2}. These two quantum phases dominate the phase diagram of $1T$-TiSe\ped{2} and are found to coexist across all the investigated H doping levels. Fully protonated H\ped{x}TiSe\ped{2} samples exhibit an impressive doping level ($\mathrm{x}\approx 2$), with negligible structural alterations, vestigial CDW signatures, and robust bulk superconductivity with possible multi-band and gapless features revealed by different experimental techniques. 
Our calculations indicate that these peculiarities of protonated TiSe\ped{2} are caused by the unique capability of H atoms to act not only as pure electron donors since, at sufficiently large H concentrations, their doping becomes band-selective and fills the Ti-$d_{z^2}$ bands, allowing the band structure of $1T$-TiSe\ped{2} to mimic that of other archetypal TMD compounds, such as the superconducting $2H$-NbSe\ped{2} or the semiconducting $2H$-MoS\ped{2}; furthermore, the hybridization between H orbitals and the Ti states at the Fermi level is found to be pivotal in reducing the electronic screening and thereby enhance the electron-phonon coupling.
Our results establish that the role played by hydrogen in SC layered materials at low pressures is crucial and yet firmly distinct from that found in high-pressure hydrides, providing a key insight for the search of high-temperature superconductivity in layered compounds at ambient pressure.
	
\section*{Results}

\begin{figure*}
	\begin{center}
		\includegraphics[keepaspectratio, width=\textwidth]{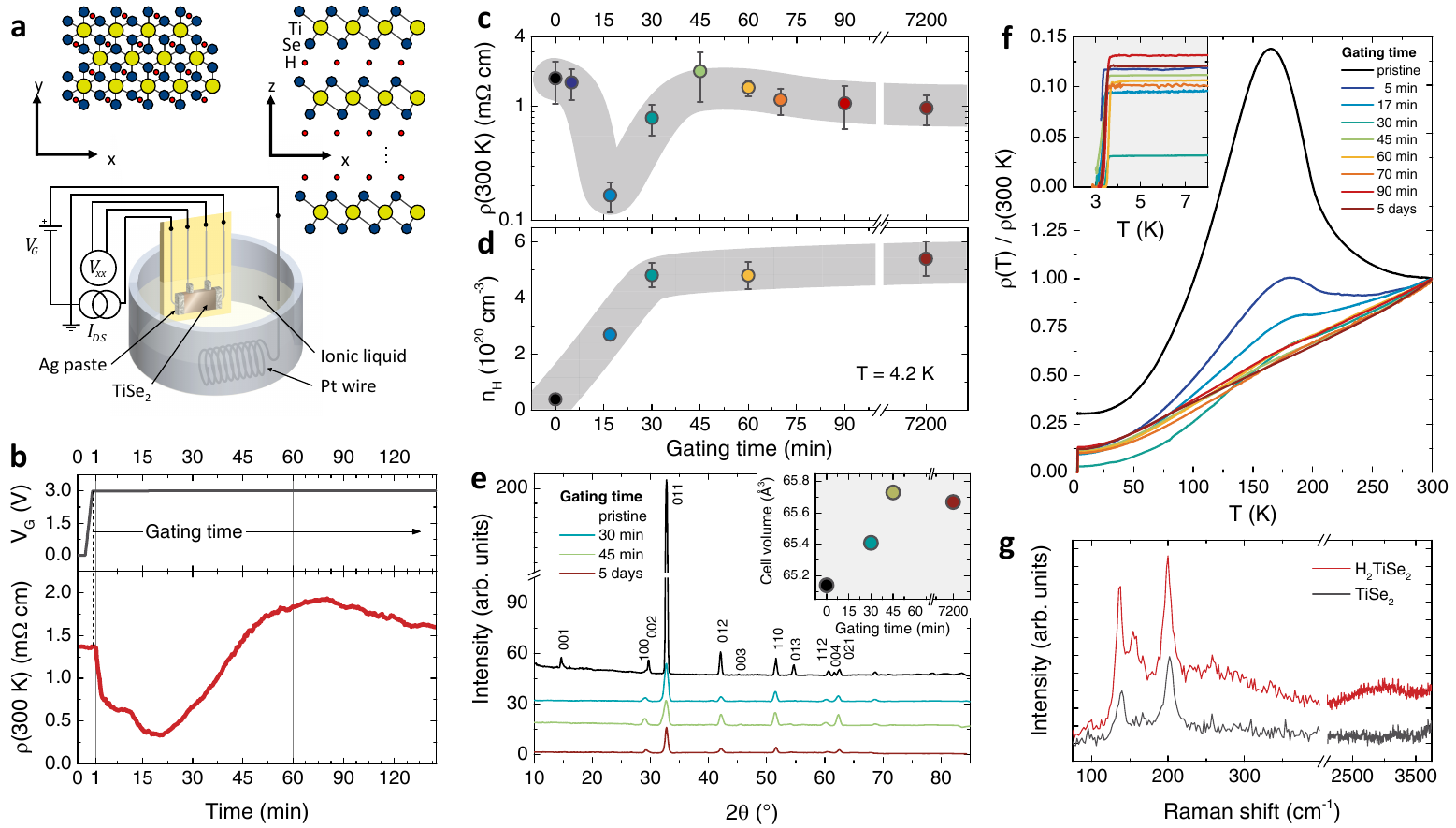}
	\end{center}
	\caption{
		\textbf{Ionic liquid gating induced protonation: electric transport and structural characterization.}
		\textbf{a}, Sketch of a TiSe\ped{2} crystal immersed in the electrochemical cell for ionic liquid gating induced protonation, including the electrical connections. The side panel shows a ball-and-stick model of the H\ped{x}TiSe\ped{2} structure with $x\lesssim 1$.
		\textbf{b}, Gate voltage $V\ped{G}$ and resistivity $\rho(300\,\mathrm{K})$ measured \textit{in situ} as a function of time, during a typical gating sequence in the electrochemical cell at ambient conditions. The gating time (i.e. the time at which the crystal is kept at $V\ped{G}=+3\,$V) is highlighted.
		\textbf{c}, $\rho(300\,\mathrm{K})$ measured \textit{ex situ} and
		\textbf{d}, Hall carrier density $n\ped{H}$ at 4.2\,K, of a series of TiSe\ped{2} crystals gated for increasing amounts of time. \add{Error bars represent the uncertainties on the measurements, which almost entirely arise from non-ideal sample geometry.} Shaded grey bands are guides to the eye.
		\textbf{e}, X-ray diffraction spectra in selected TiSe\ped{2} crystals at different gating times. The inset shows the resulting unit cell volume as a function of the gating time.
		\textbf{f}, $\rho$ as a function of temperature ($T$) for the same crystals reported in \textbf{c}, normalized by the value at 300\,K. The inset shows a magnification of the $T$ range where the superconducting transitions are observed.
		\textbf{g}, Raman spectra in ambient conditions acquired on the freshly-cleaved surfaces of a pristine TiSe\ped{2} (black curve) and a H\ped{2}TiSe\ped{2} (red curve) crystal.
	}
	\label{figure:1}
\end{figure*}

\textbf{Electric field-driven hydrogen intercalation.}
We control the H loading in our $1T$-TiSe\ped{2} samples via the ionic liquid gating-induced protonation method\,\cite{LuNature2017, CuiCPL2019, RafiqueNanoLett2019, BoeriJPCM2021}, as depicted in Fig.\,\ref{figure:1}a. When the gate voltage $V\ped{G}$ is swept in ambient conditions above a threshold voltage (Methods), the intense electric field at the TiSe\ped{2}/electrolyte interface splits the water molecules absorbed in the ionic liquid and drives the H\apex{+} ions into the van der Waals gap of the TiSe\ped{2} crystal. H incorporation in the TiSe\ped{2} lattice is confirmed by nuclear magnetic resonance (NMR) and muon spin rotation (\musr) measurements as discussed later, whereas significant intercalation by the organic ions of the ionic liquid is ruled out via X-ray photoelectron spectroscopy (XPS; see Supplementary Note 1). \textit{In-situ} monitoring of the room-temperature TiSe\ped{2} resistivity during gating (Methods) shows that $\rho(300 \mathrm{K})$ remains unaffected by the applied $V\ped{G}$ below the threshold for H intercalation, then rapidly decreases and reaches a minimum (Fig.\,\ref{figure:1}b). Maintaining $V\ped{G}=+3\,$V beyond this point leads to $\rho(300\, \mathrm{K})$ gradually increasing again over time, eventually reaching a plateau for gating times in excess of $\sim45\,$min and up to several days. 

This gate-induced change in the electric transport properties is non-volatile. Indeed, as shown in Fig.\ref{figure:1}c, \textit{ex-situ} measurements on samples gated for a fixed amount of time and then removed from the electrochemical cell give values of $\rho(300\, \mathrm{K})$ that closely follow those determined \textit{in situ}. The saturation of $\rho(300 \mathrm{K})$ for large gating times suggests that most of the H dopants are driven into the sample in the first few tens of minutes, and further increasing the gating time does not alter the average stoichiometry appreciably. This picture is confirmed by assessing the changes to the structural and electronic properties of the protonated samples. 

Low-temperature Hall effect measurements (Methods) show an increase in the Hall density $n\ped{H}$ at $T=4.2\,$K by more than an order of magnitude, going from $\sim4\cdot10^{19}$\,cm\apex{-3} in pristine TiSe\ped{2} to $\sim 5\cdot10^{20}$\,cm\apex{-3} in H\ped{2}TiSe\ped{2} (Fig.\,\ref{figure:1}d), indicating that H loading introduces a strong electron doping in the protonated TiSe\ped{2} samples. 
At the same time, a room-temperature X-ray diffraction (XRD) analysis (Fig.\,\ref{figure:1}e) shows that the H loading does not alter the $1T$ lattice structure of the pristine material, and simply expands the unit cell by up to $\sim0.9\%$. The lattice expansion is comparable to that driven by incorporation of Cu atoms in SC Cu\ped{x}TiSe\ped{2} ($\sim$\,0.6\%)\,\cite{MorosanNatPhys2006} and much smaller than the one reported upon lithiation in SC Li\ped{x}TiSe\ped{2} ($\sim$\,12\%)\,\cite{LiaoNatCommun2021}. The pronounced broadening of the XRD peaks however indicates that the protonation introduces a large degree of structural disorder in the samples, which may be partially responsible for the re-entrant behaviour of $\rho(300\,\mathrm{K})$ upon gating times in excess of those associated to the local minimum shown in Fig.\,\ref{figure:1}b,c. 
Again, both the structural and the electronic modifications to the TiSe\ped{2} crystals occur within the first $\sim 45\,$min of gating time, confirming the behaviour showcased by the room-temperature resistivity.

The dramatic effect of the protonation on the electric transport properties of TiSe\ped{2} is even more evident when these are measured as a function of temperature. Fig.\,\ref{figure:1}f shows the temperature dependence of the normalized resistivity $\rho(T)/\rho(300\,\mathrm{K})$ of TiSe\ped{2} crystals gated for different amounts of time (Methods). Pristine crystals exhibit a strongly non-monotonic $\rho(T)$ dependence with a broad peak around $\sim$\,160\,K and no sign of a SC transition. This peak is typical of 1$T$-TiSe\ped{2} and is caused by the reconstruction of the electronic band structure upon the onset of CDW order, which shifts the Fermi level and partially gaps the Fermi surface\,\cite{MorosanNatPhys2006, LiNature2016, MorosanPRB2010, WuPRB2007, WangAPL2018}. 
\tcb{The monotonically-increasing $\rho(T)$ dependence up to $T\ped{CDW}$ is also commonly observed in high-quality 1$T$-TiSe\ped{2} crystals, but whether it originates from a semimetallic band overlap\,\cite{DiSalvoPRB1976, JaouenPRB2019} or from a narrow semiconducting gap in the presence of a small extrinsic electron doping\,\cite{HuangPRB2017, CampbellPRM2019, WatsonPRB2019} remains currently uncertain.}
A small gating time of $\sim 5\,$min strongly suppresses the intensity of the CDW peak and leads to the appearance of SC with an onset transition temperature $T\ped{c}\apex{on}\simeq3.3\,$K. Larger gating times up to $\sim45\,$min lead to a further reduction in the intensity of the CDW peak and simultaneously slightly increase $T\ped{c}\apex{on}$ to $\sim3.7\,$K (with no clear dependence, see inset to Fig.\,\ref{figure:1}f). Also here, gating times in excess of $\sim 45\,$min lead to a clusterization of the $\rho(T)$ curves around a common behaviour, characterized by a complete SC transition with $T\ped{c}\apex{on}\sim3.7\,$K and a faint trace of CDW order in the form of a slope change below $\sim 200\,$K.
\tcb{Consistent with the XRD results, a large degree of structural disorder in these highly-doped samples is evidenced by the four-fold reduction in their residual resistivity ratio (Methods), from its maximum value $\sim$\,32 attained at a gating time of $\sim$\,30\,min, to a final value $\sim$\,8 at a gating time of $5$\,days.}

The number of intercalated protons in
H\ped{x}TiSe\ped{2} samples pertaining to this state of clusterized $\rho(T)$ and saturated $n\ped{H}$ and cell volume (see Fig.\,\ref{figure:1}f, d and e respectively) is quantified by means of $^{1}$H-NMR (Methods and Supplementary Note 2). In particular, the decay of the spin-echo amplitude is measured both in a collection of fully-doped TiSe$_{2}$ crystals and in a reference sample of hexamethylbenzene (C$_{12}$H$_{18}$). A comparison between the signal amplitudes for the two samples shows that the intercalated TiSe$_{2}$ crystals contain an average number $\mathrm{x} = 2.0 \pm 0.3$ of protons per TiSe$_{2}$ formula unit.
\tcb{Assuming the H concentration to be linearly proportional to $n\ped{H}$ at $T=4.2\,$K, this then allows estimating an average $\mathrm{x} \sim 0.4$ for intercalated TiSe\ped{2} to exhibit an incomplete resistive transition with reduced $T\ped{c}\apex{on}$ (gating time of $\sim 5$\,min), and an average $\mathrm{x} \sim 1$ for a complete resistive transition to develop (gating time of $\sim 17$\,min).}

H loading also affects the vibrational properties of TiSe\ped{2}, as demonstrated by room-temperature Raman spectroscopy on freshly-cleaved surfaces (Fig.\,\ref{figure:1}g; see Methods). Spectra acquired on pristine TiSe\ped{2} exhibit the two main Raman modes of 1$T$-TiSe\ped{2}, the E\ped{g} mode at $\sim$\,138\,cm\apex{-1} and the A\ped{1g} mode at $\sim$\,202\,cm\apex{-1}, consistent with the literature\,\cite{LiAPL2016, DuongACSNano2017, HolyPRB1977, SugaiSSC1980, UchidaPBC1981}, and no additional peaks. In spectra acquired on H\ped{2}TiSe\ped{2}, the same E\ped{g} and the A\ped{1g} peaks are redshifted to $\sim$\,136\,cm\apex{-1} and $\sim$\,199\,cm\apex{-1} respectively. Two additional, broader peaks emerge in H\ped{2}TiSe\ped{2} at $\sim$\,156\,cm\apex{-1} and $\sim$\,260\,cm\apex{-1}, which could be ascribed to intercalant-activated two-phonon processes at the L and/or M points\,\cite{DuongACSNano2017, HolyPRB1977, SugaiSSC1980, JaswalPRB1979}\tcb{, and a very broad band appears at high wavenumbers centered around $\sim$\,2900\,cm\apex{-1}.
Our first-principle calculations (see Supplementary Note 3) indicate that} the appearance of the peaks \tcb{at $\sim$\,156\,cm\apex{-1} and $\sim$\,260\,cm\apex{-1}} can be justified by the removal of symmetry of the 1$T$-TiSe\ped{2} system resulting from the incorporation of H atoms \tcb{in both atomic (H\ped{1}TiSe\ped{2}) and molecular (H\ped{2}TiSe\ped{2}) forms. The broad band centered around $\sim$\,2900\,cm\apex{-1} can instead be naturally interpreted as originating from H\ped{2} molecules confined in the TiSe\ped{2} matrix, with a renormalized phonon frequency\,\cite{FuteraJPCC2017, OkamotoPRB1997}. The simultaneous observation of these multiple peaks is therefore evidence of a strong disorder in the H concentration, phase and distribution in the TiSe\ped{2} matrix.} 
In this case, the observation of consistent Raman spectra across different spots in the H\ped{2}TiSe\ped{2} crystals could be interpreted as the phase separation between regions with different H concentration to occur over length scales smaller than the laser spot size ($\sim1\,\upmu$m).

\begin{figure*}
	\begin{center}
			\includegraphics[keepaspectratio, width=\textwidth]{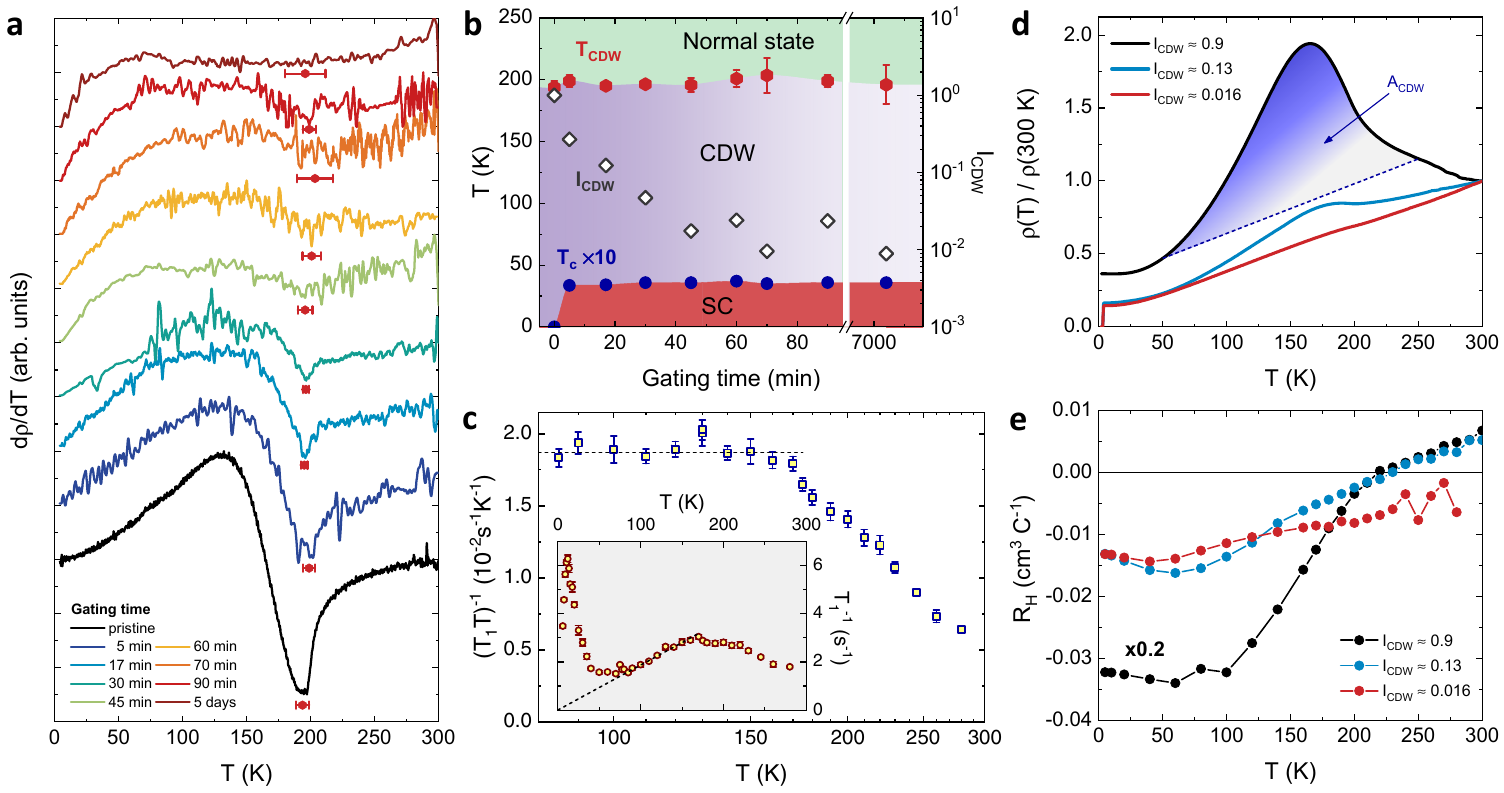}
	\end{center}
	\caption{
		\textbf{Charge density waves in H\ped{x}TiSe\ped{2}.}
		\textbf{a}, First derivative of the resistivity $\rho$ as a function of temperature $T$, $d\rho/dT$, obtained by numerical derivation of the $\rho(T)$ curves shown in Fig.\,\ref{figure:1}f for increasing gating time. Red bars highlight the position of the local minimum associated with the onset of CDW order \add{and the uncertainty on its determination}.
		\textbf{b}, Temperature-gating time phase diagram of H\ped{x}TiSe\ped{2}. The onset temperature of CDW order, $T\ped{CDW}$, and the superconducting transition temperature, $T\ped{c}$, are plotted as a function of gating time (left scale). The CDW intensity $I\ped{CDW}$ (right scale) is calculated as $A\ped{CDW}(t)/A\ped{CDW}(0)$, where $A\ped{CDW}(t)$ is the area contained between each normalized resistivity curve, $\rho(T)/\rho(300\,\mathrm{K})$, and a straight line connecting the data points at $50$ and $250\,$K. Shaded areas highlight the different phases observed in the system.
		\textbf{c}, $T$ dependence of the spin-lattice relaxation rate \Tone\ at $\mu_{0}H \simeq 3.5$\,T (inset). The dashed line is a linear fit to the experimental data highlighting a well-defined Korringa-like behaviour for $T \lesssim 170$ K and a marked decrease of \Tone\ upon increasing temperature above $210$ K. The main panel reports the $T$ dependence of $\left(T_1T\right)^{-1}$, making the crossover between the different behaviours even clearer. The dashed line is a constant line corresponding to the fitting curve reported in the inset. \add{Error bars represent the uncertainties on the measurements.}
		\textbf{d}, Normalized resistivity and
		\textbf{e}, Hall coefficient $R\ped{H}$ as a function of $T$ in three TiSe\ped{2} crystals with different $I\ped{CDW}$. Both protonated crystals (intermediate and low $I\ped{CDW}$) are superconducting with $T\ped{c}\apex{on}\approx 3.7\,$K, which is defined as the threshold where $\rho(T)$ reaches 95\% of its normal-state value. The shaded blue region in \textbf{d} highlights the area from which $A\ped{CDW}(t)$ is calculated.
	}
	\label{figure:2}
\end{figure*}

\textbf{Charge density waves in H\ped{x}TiSe\ped{2}.}
The evolution of the CDW order with increasing gating time $t$ can be more accurately tracked by considering its onset temperature, $T\ped{CDW}$, and the intensity of the CDW peak, $I\ped{CDW}$. As firmly established in the literature\,\cite{DiSalvoPRB1976, LiNature2016, LiaoNatCommun2021}, the former can be identified with the position of the inflection point in the $\rho(T)$ curves, that corresponds to the minimum in $d\rho/dT$. As shown in Fig.\,\ref{figure:2}a, which displays $d\rho/dT$ as a function of $T$, the visibility of the local minimum is suppressed by increasing gating time, but its position remains basically unchanged and close to its pristine value $T\ped{CDW}\approx198\,$K. The intensity can be evaluated as $I\ped{CDW}=A\ped{CDW}(t)/A\ped{CDW}(0)$, where $A\ped{CDW}(t)$ is the area contained between each normalized $\rho(T)$ curve (shown in Fig.\,\ref{figure:1}f) and a line connecting its two values at 50 and 250\,K\,\cite{LiaoNatCommun2021}, as exemplified in Fig.\,\ref{figure:2}d. The trends of $T\ped{CDW}$ and $I\ped{CDW}$ as a function of gating time are summarized in Fig.\,\ref{figure:2}b (filled red hexagons and hollow black diamonds, respectively). Clearly, while $T\ped{CDW}$ is almost constant, $I\ped{CDW}$ is rapidly suppressed upon increasing gating time, and becomes of the order of 1\,\% of its pristine value in H\ped{2}TiSe\ped{2} samples. This behaviour is similar to that reported in Li\ped{x}TiSe\ped{2} crystals\,\cite{LiaoNatCommun2021}, but differs from that of Cu\ped{x}TiSe\ped{2}\,\cite{MorosanNatPhys2006, ZhaoPRL2007}, Ti\ped{1+x}Se\ped{2}\,\cite{JaouenPRB2019}, electrostatically ion-gated TiSe\ped{2}\,\cite{LiNature2016}, and TiSe\ped{2} under pressure\,\cite{KusmartsevaPRL2009}, where the suppression of $I\ped{CDW}$ was accompanied by a decrease of $T\ped{CDW}$. 
The insensitivity of $T\ped{CDW}$ to increasing doping was interpreted, in the case of Li\ped{x}TiSe\ped{2}, as a lack of full in-plane percolation of the dopants, leading to a spatial separation between SC Li-rich regions and CDW-ordered pristine regions\,\cite{LiaoNatCommun2021}. 
The pronounced increase in structural disorder detected by both XRD and Raman spectroscopy in our samples raises the question of whether a similar picture may hold also in the case of H\ped{x}TiSe\ped{2}.

In general, however, an explanation relying purely on phase separation \tcb{between H-deficient, CDW-ordered regions and H-rich, SC regions} might be too simplistic, since it is not straightforward that the presence of dopants destroys the CDW order in TiSe\ped{2} compounds. For example, scanning tunnelling microscopy measurements in Cu\ped{x}TiSe\ped{2} indicate that CDW order robustly survives even in Cu-rich areas where the local electron doping is large\,\cite{SperaPRB2019, SperaPRL2020}. 
On the other hand, disorder itself is able to suppress the signatures of CDW order in electric transport, as was reported in the case of irradiated NbSe\ped{2}\,\cite{ChoNatCommun2018}.
In this context, the temperature dependence of the Hall coefficient $R\ped{H}$ (Methods and Supplementary Note 4) suggests that a weakened CDW order could survive in dopant-rich areas also in the case of H\ped{x}TiSe\ped{2}, as shown in Fig.\,\ref{figure:2}d,e. In agreement with the established literature\,\cite{DiSalvoPRB1976, WangAPL2018, LiaoNatCommun2021, KnowlesPRL2020}, pristine TiSe\ped{2} exhibits both a strong CDW peak in $\rho(T)$ (black curve in Fig.\,\ref{figure:2}d, $I\ped{CDW}\approx 0.9$) and a change of sign in $R\ped{H}$ around $T\approx 215\,$K (black curve in Fig.\,\ref{figure:2}e). This occurs when the onset of CDW order opens a gap at the Fermi level in part of the band structure, causing a reconstruction in the Fermi surface \tcb{which changes the dominant character of the charge carriers from holonic for $T \gtrsim T\ped{CDW}$ to electronic for $T \lesssim T\ped{CDW}$}\,\cite{DiSalvoPRB1976, WangAPL2018, LiaoNatCommun2021, KnowlesPRL2020}. 
An intermediate H doping (blue curve in Fig.\,\ref{figure:2}d,e) suppresses the CDW peak in $\rho(T)$ ($I\ped{CDW}\approx 0.13$) and strongly reduces the absolute value of $R\ped{H}$, but a sign change is still clearly present at $T\approx 225\,$K, indicating that the Fermi surface still undergoes a reconstruction.
H\ped{2}TiSe\ped{2} alone (red curve in Fig.\,\ref{figure:2}d,e) exhibits no sign change in $R\ped{H}$ up to $\sim 300\,$K, and a CDW peak in $\rho(T)$ reduced to a slope change ($I\ped{CDW}\approx 0.016$). 
\tcb{However, the Hall coefficient remains strongly $T$-dependent, resulting in a tenfold reduction in $n\ped{H}$ as the sample is cooled from 300\,K to 4\,K. This freeze-out of the free charge carriers can be ascribed to the presence of trap states introduced by structural disorder, but could also be assisted by a residual partial gapping of the Fermi surface if a weakened CDW order survives in the H-rich regions.
The absence of a sign change in $R\ped{H}$} could indicate either that the CDW gap still opens around $\sim 200\,$K in H\ped{2}TiSe\ped{2} but is partially shifted below the Fermi level by the increasing electron doping (similar to the case of Cu\ped{x}TiSe\ped{2}\,\cite{SperaPRB2019, SperaPRL2020}), or that the disorder introduced by the H intercalation suppresses long-range CDW order in the H-rich regions and allows only short-range CDW order to survive (as in the case of irradiated NbSe\ped{2}\,\cite{ChoNatCommun2018}). A combination of the two effects is of course also possible.
\begin{figure*}
	\begin{center}
		\includegraphics[keepaspectratio, width=\textwidth]{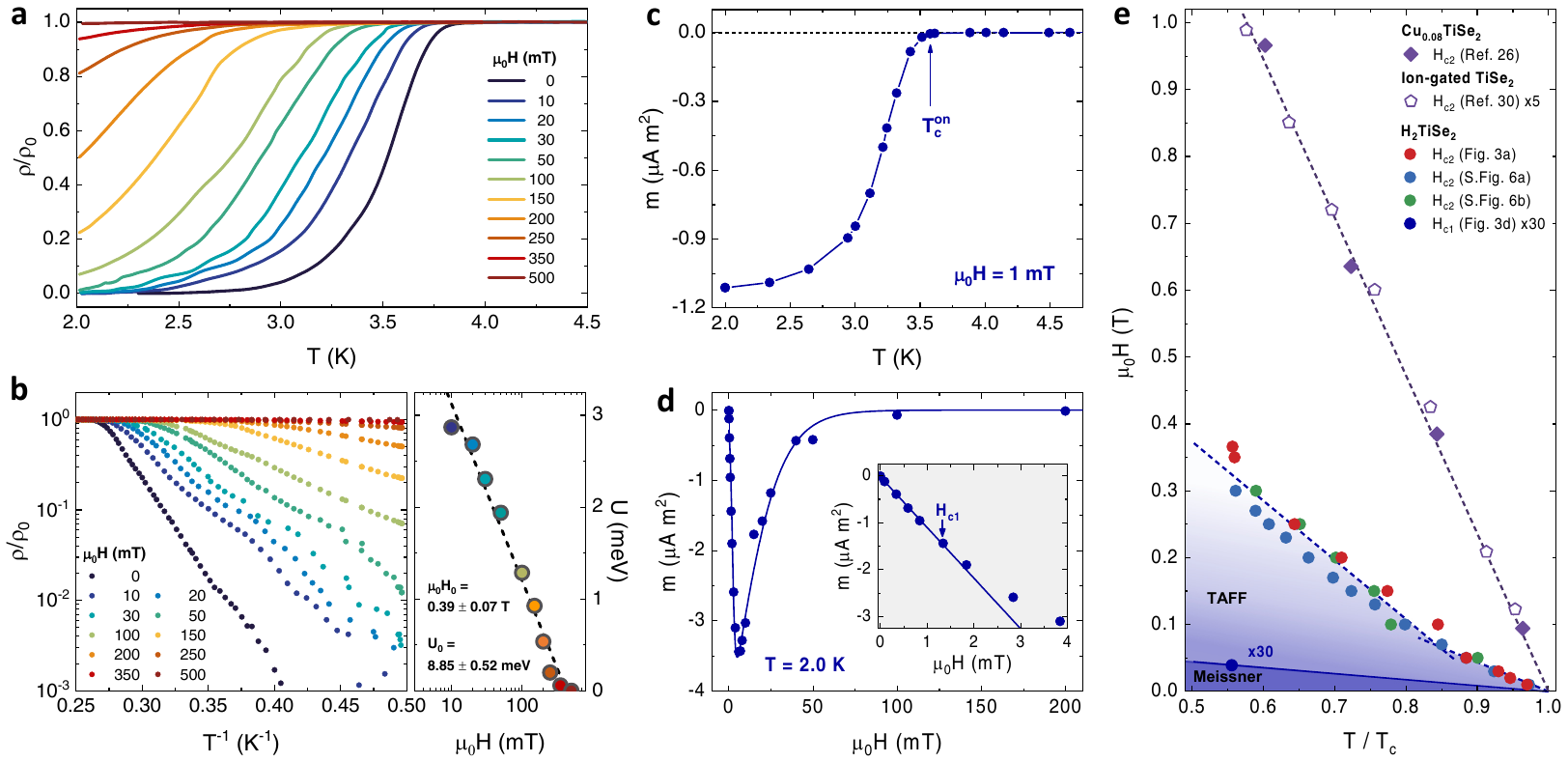}
	\end{center}
	\caption{
		\textbf{Magnetotransport and magnetization of superconducting H\ped{2}TiSe\ped{2}.}
		\textbf{a}, Resistivity $\rho$ of a H\ped{2}TiSe\ped{2} crystal as a function of temperature $T$ in the superconducting transition, normalized by its normal-state value $\rho_0$, for increasing magnetic field $H$ applied orthogonal to the crystal $ab$ plane.
		\textbf{b}, Same data of \textbf{a} plotted in semilogarithmic scale as a function of $T^{-1}$, highlighting the Arrhenius behaviour characteristic of thermally-activated flux flow (TAFF). The side panel shows the TAFF activation energy $U$ as a function of $\mu_0H$. Dashed black line is the linear fit to the data in semilogarithmic scale which allows determining the characteristic energy $U_0$ and magnetic field $\mu_0H_0$.
		\textbf{c}, $T$-dependence of the magnetic moment $m$,  
		recorded under zero-field cooling (ZFC) conditions with $\mu_0H = 1$\,mT magnetic field applied perpendicular to the $ab$ plane. The arrow indicates the $T$ below which the magnetic moment starts deviating from its normal-state value, thus defining the diamagnetic onset temperature $T\ped{c}\apex{on} = 3.6 \pm 0.1$\,K.
		\textbf{d}, ZFC magnetic moment as a function of $H$ measured isothermally at $T = 2$\,K. The line is a guide to the eyes. The inset shows a magnification of the low-field region, with a linear fit on the first six points (line). The arrow indicates the field above which the magnetic moment starts deviating from a linear dependence, thus defining the first critical field $H\ped{c1} = 1.3 \pm 0.3$\,mT.
		\textbf{e}, $H-T$ phase diagram of H\ped{2}TiSe\ped{2} constructed by combining the results from resistivity and magnetization measurements on multiple different crystals. Shaded blue regions highlight the Meissner state and the TAFF state. Red and indigo circles are obtained from the data shown in \textbf{a} and \textbf{d} respectively. Blue and green circles are obtained from Supplementary Fig.\,6a and b respectively. Dashed blue lines are linear fits to the resistive critical magnetic field $H\ped{c2}$ (defined as the threshold for reaching $\rho(T)=0.95\rho_0$) above and below the kink at $T/T\ped{c}\approx0.85$. The resistive $H\ped{c2}$ data for Cu\ped{x}TiSe\ped{2}\,\cite{MorosanNatPhys2006} and ion-gated TiSe\ped{2}\,\cite{LiNature2016} are shown for comparison.
	}
	\label{figure:3}
\end{figure*}

Strong indications of a reconstruction of the Fermi surface around $T \sim 200$\,K are also {\color{blue}obtained from NMR measurements of the {}$^{1}$H spin-lattice relaxation rate \Tone\ in H\ped{2}TiSe\ped{2} samples at $\mu_{0}H \simeq 3.5$\,T (Methods). As shown in the inset to Fig.\,\ref{figure:2}c, \Tone\ exhibits a linear dependence on $T$ for $80$\,K $\lesssim T \lesssim 170$\,K consistent with the so-called Korringa behaviour proper of relaxation processes mediated by itinerant electrons\,\cite{Slichter}. This is in agreement with the results of NMR measurements in other TMDs such as VSe$_{2}$, VS$_{2}$, and IrTe$_{2}$ within the CDW phase\,\cite{Tsu81,Tsu83,Miz02}}. Higher temperatures $T \gtrsim 170$\,K mark a clear departure from the linear-in-temperature trend. In particular, a plateau of \Tone\ between $170$\,K and $210$\,K is followed by a suppression of \Tone\ upon increasing $T$. The crossover between the two different behaviours is even clearer in the $T$ dependence of $\left(T_1T\right)^{-1}$ (main panel of Fig.\,\ref{figure:2}c) {\color{blue}and it seemingly correlates with the onset of the CDW phase. In the ideal case of a Fermi gas, it is well-known that $\left(T_1T\right)^{-1} \propto \varrho(E_{F})$ -- i.e., the density of states at the Fermi energy\,\cite{Slichter}. The observed drop of $\left(T_1T\right)^{-1}$ for $T \gtrsim 170$\,K is then strongly indicative of a progressive decrease of $\varrho(E_{F})$ upon increasing temperature above the critical temperature of the CDW phase, at variance with what is observed in VSe$_{2}$, VS$_{2}$, and IrTe$_{2}$\,\cite{Tsu81,Tsu83,Miz02}. This peculiar behaviour will be discussed in detail elsewhere\,\cite{Pra23}. Finally, we} stress that the sharp maximum observed in the spin-lattice relaxation rate at around $10$ K has already been reported in pristine TiSe$_{2}$\,\cite{Dupree1977}. This observation confirms that the intercalated {}$^{1}$H nuclei probe the intrinsic electronic properties of the host material.

\textbf{Superconductivity in H\ped{x}TiSe\ped{2}.}
The evolution of the SC phase in H\ped{x}TiSe\ped{2} appears completely unaffected by the presence of CDW order, as summarized by the values of $T\ped{c}$ as a function of gating time (blue circles in Fig.\,\ref{figure:2}b). A complete SC transition with a sizeable $T\ped{c}\apex{on}\sim3.7\,$K is observed not only in the fully-doped 
H\ped{2}TiSe\ped{2}, where no Fermi-surface reconstruction is detected by Hall effect, but also in the partially-doped H\ped{x}TiSe\ped{2}, where the Fermi surface is certainly partially gapped (see inset in Fig.\,\ref{figure:1}f). In a magnetic field $\mu_0H$ perpendicular to the $ab$ plane of the H\ped{2}TiSe\ped{2} crystal (Fig.\,\ref{figure:3}a and Supplementary Note 5; see Methods) the SC transition broadens and shifts to lower $T$. At $2\,$K, a well-defined zero-resistance state (ZRS) is observed at least up to $\mu_0H = 30\,$mT and the intermediate dissipative regime exhibits the typical behaviour of a thermally-activated flux flow (TAFF), $\rho(T,H) = \rho_0 \exp[-U(H)/k\ped{B}T]$, at any $T$ and $H$. This is clear from Fig.\,\ref{figure:3}b, where $\rho/\rho_0$ is plotted in logarithmic scale against $T^{-1}$. 
The activation barrier for vortex motion, obtained from these curves, scales logarithmically with increasing magnetic field, $U(H) = U_0 \ln(B_0/\mu_0H)$ (right panel in Fig.\,\ref{figure:3}b), indicating a collective flux creeping\,\cite{BlatterRMP1994}. A fit to the $U(H)$ data gives $U_0\approx8.9\,$meV and $B_0\approx0.39\,$T, comparable to the values found in ion-gated TiSe\ped{2}\,\cite{LiNanoLett2019}.
The observation of this direct transition from ZRS to TAFF already at relatively large $T\gtrsim 2$\,K is analogous to those found in overdoped Li\ped{x}TiSe\ped{2}\,\cite{LiaoNatCommun2021} and Cu\ped{x}TiSe\ped{2}\,\cite{MorosanNatPhys2006}, and provides no evidence that H\ped{x}TiSe\ped{2} may host the anomalous quantum metallic behavior that was claimed to destroy the ZRS at arbitrarily small magnetic fields in ion-gated TiSe\ped{2}\,\cite{LiNanoLett2019} and in under- and optimally-doped Li\ped{x}TiSe\ped{2}\,\cite{LiaoNatCommun2021}.

Fig.\,\ref{figure:3}c shows the $T$-dependence of the zero-field cooled (ZFC) magnetic moment $m(T)$ measured below $\sim 5\,$K at $\mu_0 H = 1$\,mT (Methods). A rather sharp transition is observed at $T\ped{c}\apex{on} = 3.6\pm0.1$\,K, in reasonable agreement with the $T_{c}$ value determined by transport measurements. 
As expected for the Meissner state, the isothermal $m(H)$, measured at $T=2\,$K following a ZFC procedure, decreases linearly with increasing $H$, up to $1.3\pm0.3$\,mT (Fig.\,\ref{figure:3}d). Above this field, $m(H)$ starts deviating from a linear dependence, thus defining the lower critical field $H\ped{c1}(2\,\mathrm{K})$. Upon further increasing $H$, $m(H)$ first reaches a minimum at $\approx5.8$\,mT and then it decreases in modulus, returning to zero at the upper critical field $\gtrsim 100$\,mT, in fairly good agreement with the $H_{c2}(T)$ value determined from transport measurements (where we use the criterion of 95\% normal-state resistivity).

By combining the resistivity- and magnetization measurement results, we can construct the SC $T{-}H$ phase diagram of H\ped{2}TiSe\ped{2} shown in Fig.\,\ref{figure:3}e. The Meissner state exists for magnetic fields $H\lesssim H\ped{c1}$, above which SC vortices penetrate the sample volume and move in the TAFF regime until $H=H_{c2}$, here defined as the field where $\rho$ reaches 95\% of its normal-state value. The $T$-dependence of $H_{c2}$ shows a prominent kink around $\sim0.85T_{c}$, where its slope changes abruptly from $\approx -0.23$ to $\approx -0.13$\,T/K. These two slopes correspond to an in-plane coherence length $\xi_{ab}(0)$ of 26 and 20\,nm, respectively (Methods). We note that, as shown in Fig.\,\ref{figure:3}e, this behavior is invariably observed across all the measured H$_2$TiSe$_2$ samples with little sample-to-sample variation, even in those that exhibit a strongly biphasic resistive transition (Supplementary Fig.\,6a) or a weakly-localized normal state (Supplementary Fig.\,6b). This indicates that it is mostly insensitive to disorder and sample inhomogeneities, and therefore is likely an intrinsic property of the material. As a consequence, since this feature deviates considerably from the single-band mean-field behaviour exhibited by both the 3D Cu\ped{x}TiSe\ped{2} and the ion-gated 2D TiSe$_{2}$, it suggests that H\ped{x}TiSe\ped{2} may be a multi-band superconductor, where it is instead commonly observed\,\cite{XingSciRep2017, GurevichPRB2003, DingNL2022}.

\begin{figure}
	\begin{center}
			\includegraphics[keepaspectratio, width=\columnwidth]{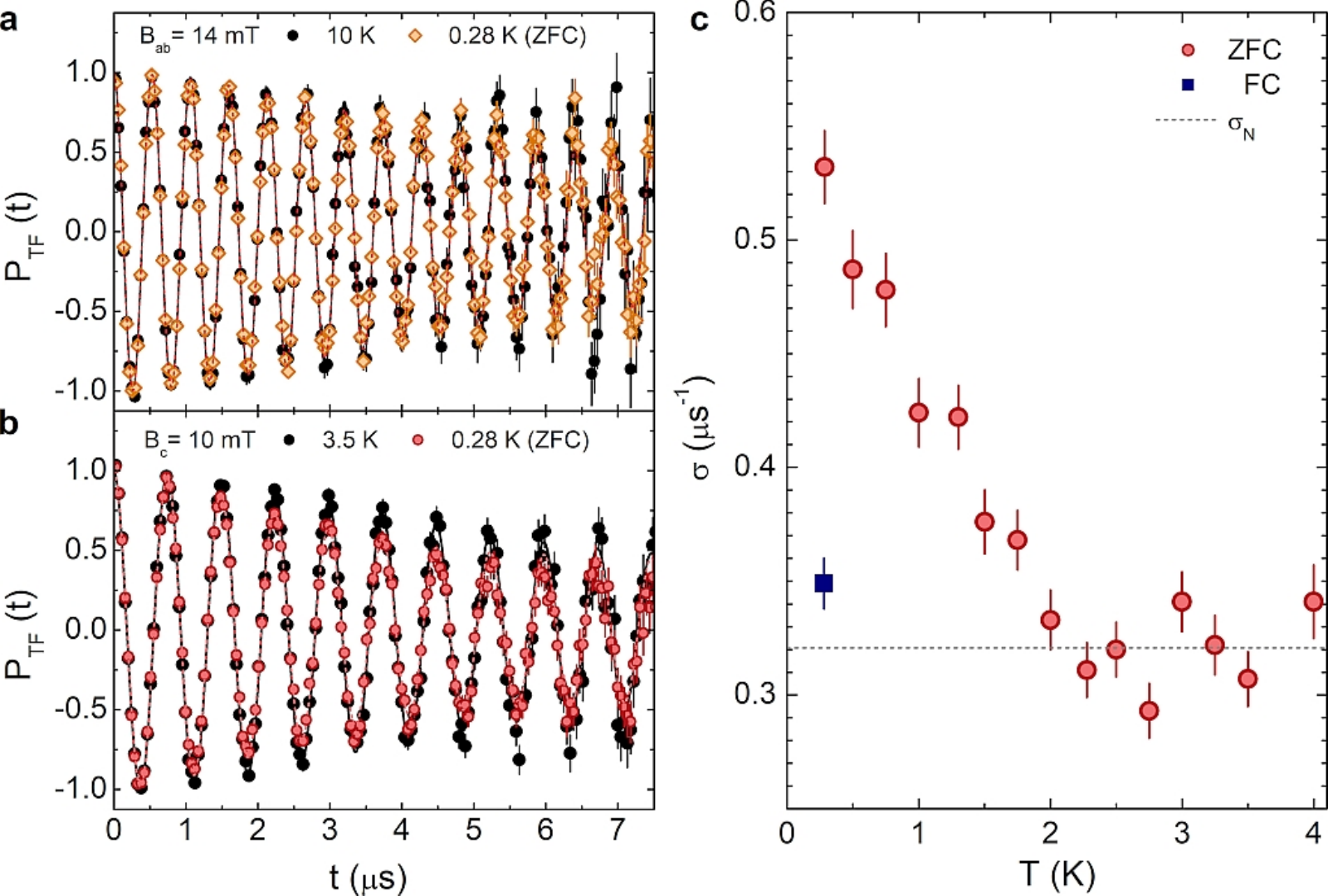}
	\end{center}
	\caption{\label{figure:4} 
	\textbf{Muon spin rotation in superconducting H\ped{2}TiSe\ped{2}.}
	 \textbf{a},\textbf{b}, Zero-field cooling transverse-field muon-spin polarization $P_\mathrm{TF}(t)$ in
	 a 14-mT magnetic field applied parallel to the $ab$ planes, (\textbf{a}), or in a 10-mT field applied parallel to the $c$-axis, (\textbf{b}). Solid lines are fits to the model described by Eq.\,\eqref{eq:pt}.
	 \textbf{c}, Temperature dependence of the depolarization rate (red circles) in a magnetic field $\mu_{0}H= 10$\,mT, applied parallel to the $c$ axis. Note the lack of saturation at low temperatures. The dark-blue square represents the depolarization rate recorded after field cooling in the same applied magnetic field. The grey dashed line represents the average nuclear dipolar contribution, here, $\sigma\ped{n}=0.32\pm0.02$\,$\upmu$s$^{-1}$. \textcolor{blue}{Error bars represent the uncertainties on the measurements (\textbf{a-b}) and on the fits (\textbf{c}).}
	}
\end{figure}

Further insight into the SC properties of the H\ped{2}TiSe\ped{2} is obtained by transverse-field (TF) \musr\ measurements\,\cite{Blundell1999, Amato1997} (Methods and Supplementary Note 6). Figs.\,\ref{figure:4}a,b show the zero-field cooling (ZFC) TF-polarization curves recorded either in an applied field parallel to the $ab$-planes, $\mu_{0}H = 14$\,mT, or parallel to the $c$-axis, $\mu_{0}H = 10$\,mT. As to the former (panel a), no significant differences were detected between datasets taken at $T=0.28$\,K and $T>T\ped{c}$. For fields applied parallel to the $c$-axis, instead, clearly distinct datasets are observed in the SC- and the normal state. 
The volume fraction of those muons probing the vortex state, $f\ped{sc}$, was determined by fitting the time-dependent TF polarization data by using a two-step calibration procedure [see details in Methods and Eq.\,\eqref{eq:pt}]. We find a lower limit for the SC volume fraction $f\ped{sc}=44\pm1$\%, thus confirming the bulk nature of H\ped{x}TiSe\ped{2} superconductivity, leaving however open the possibility that some sample regions may still be normal, as expected in case of an inhomogeneous distribution of H dopants.

In the ZFC case, the temperature dependence of the 
in-plane Gaussian depolarization rate of the implanted muons 
$\sigma(T)$, resulting from fits to Eq.\,\eqref{eq:pt} (Methods), is shown in Fig.\,\ref{figure:4}c. It includes the contributions of $\sigma\ped{n}$ and $\sigma\ped{sc}$, the depolarization rates in the normal and SC state, respectively, and is given by $\sigma=\sqrt{\sigma\ped{n}^{2}+\sigma\ped{sc}^{2}}$. 
$\sigma\ped{n}$ is due to the contribution of the randomly-oriented nuclear dipole moments, here fully dominated by the intercalated H nuclei, whose moments are only partially quenched by the applied magnetic field. Therefore, the observation of a finite $\sigma\ped{n}=0.32\pm0.02$\,$\upmu$s$^{-1}$in the normal state (dashed line in Fig.~\ref{figure:4}c) provides an independent confirmation of the sizeable H intake in the H\ped{2}TiSe\ped{2} samples.
The effective magnetic penetration depth was estimated by measuring also a TF spectrum at base $T$ (dark-blue square in Fig.\,\ref{figure:4}c), now using a standard field-cooling (FC) procedure, which provides a SC-depolarization rate $\sigma_\mathrm{sc}= 0.14\pm0.03$\,$\upmu$s$^{-1}$ at $T=0.28$\,K. This, in turn, implies a rather large value for the in-plane magnetic penetration depth, $\lambda_{ab}= 900 \pm 100$\,nm (Methods), indicative of a low superfluid density.
Furthermore, the $T$ dependence of $\sigma$ does not show any signs of saturation down to the lowest temperature (0.28\,K). At the same time, auxiliary zero-field (ZF) measurements confirm that time-reversal symmetry is preserved in H\ped{2}TiSe\ped{2} (see Supplementary Note 7). This result suggests that H\ped{2}TiSe\ped{2} hosts a gapless-like SC state, thus making future spectroscopic investigations of the symmetry of SC gap(s) of this material highly desirable.

\textbf{Density functional theory of H\ped{x}TiSe\ped{2}.}
\begin{figure*}
	\begin{center}
		\includegraphics[keepaspectratio, width=0.8\textwidth]{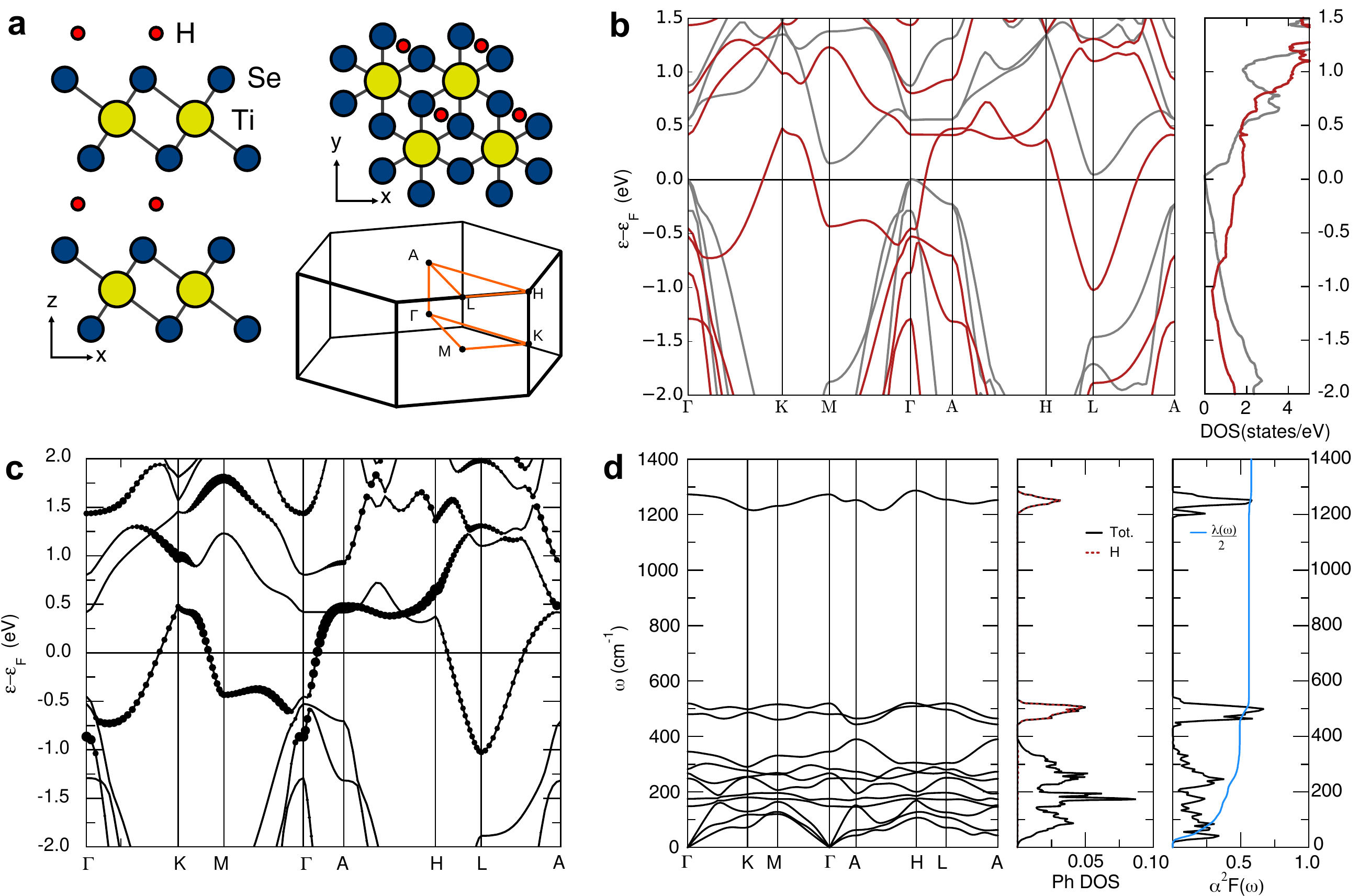}
	\end{center}
	\caption{
		\textbf{First-principles electronic and phononic structures of H\ped{x}TiSe\ped{2}.}
		\textbf{a}, Ball-and-stick model of the protonated $1T$-TiSe\ped{2} structure in the ground-state adsorption site with one H atom placed in the van der Waals gap at the Se-Se bridge position (also shown in Fig.\,\ref{figure:1}a). The Brillouin zone with high symmetry points is also shown.
		\textbf{b}, Electronic band structure and density of states of $1T$-H\ped{1}TiSe\ped{2} in the structure shown in panel \textbf{a} (red) compared with the pristine $1T$-TiSe\ped{2} dispersion (gray).
            \textbf{c}, Orbital character of the electronic band structure of $1T$-H$_1$TiSe$_2$, with the $d_{z^2}$ character of the bands indicated by the size of the black circles.
		\textbf{d}, Phonon dispersion relations (left panel), density of states (central panel), and electron-phonon spectral function $\alpha^2F(\omega)$ (right panel) of pressurized $1T$-H$_1$TiSe$_2$. The spectral contributions of the H-derived modes are highlighted as red dashed lines. The right panel also reports the frequency-dependence of the total electron-phonon coupling $\lambda(\omega)$ as a solid blue line.}
	
	\label{figure:theory}
\end{figure*}
The effects of H intercalation on the structural, electronic and SC phase of 1$T$-H\ped{x}TiSe$_2$ were investigated from a theoretical standpoint by means of \textit{ab initio} density functional theory (DFT) calculations (Methods and Supplementary Note 8). 
Simulations based on a random-search algorithm reveal the existence of several metastable intercalation sites for H in the 1$T$-TiSe\ped{2} lattice, indicating that it can easily be trapped (at $T=0$) in different local energy minima.
In particular, for concentrations up to $\mathrm{x}\lesssim 1$ H preferentially sits in the van der Waals gap at the bridge position between two Se atoms belonging to different planes (Fig.\,\ref{figure:theory}a).

From the electronic standpoint, for $\mathrm{x}\lesssim 0.1$ H acts as an electron donor, rigidly shifting the Fermi level in the conduction band of TiSe\ped{2} where the electronic states exhibit a predominant in-plane character (Supplementary Fig.\,10a). 
The simple rigid-band-shift picture breaks down at intermediate concentrations ($\mathrm{x}\approx 1$), where H doping leads to a strong deformation of the intrinsic electronic dispersion of $1T$-TiSe$_2$ (shown as the grey curve in Fig.\,\ref{figure:theory}b), induces a sizeable hybridization between the TiSe$_2$ bands and the H-derived orbitals, and completely switches the orbital character of the bands which cross the Fermi level (Fig.\,\ref{figure:theory}b, red curve). Indeed, the positive potential induced by the ionized H atoms in the van der Waals gap of $1T$-TiSe$_2$ lowers the energy of the bands with orbital character extending in the inter-layer gap, like the $d_{z^2}$ band (the intensity of which is depicted as solid circles in Fig.\,\ref{figure:theory}c). This band structure engineering is crucial for the development of SC order, since the $d_{z^2}$ character of the states at the Fermi energy was already found in the SC TMD compounds NbS$_2$ and NbSe$_2$, where the $d_{z^2}$ bands are partially-filled\,\cite{10.1088/2053-1583/ab23c0, PhysRevB.92.134510, PhysRevLett.119.087003}.

Conversely, a further increase in doping level, to the concentration determined in fully-doped H\ped{x}TiSe\ped{2} samples by NMR ($\mathrm{x}\approx 2$), leads to the emergence of two possible phases. One is characterized by the formation of H$_2$ molecules in the van der Waals gap and a band structure akin to that of intrinsic 1$T$-TiSe\ped{2} (Supplementary Fig.\,10b). The other is a metastable insulating phase, achieved when the stoichiometry is exactly 2 H atoms per unit cell (Supplementary Fig.\,10d).
The actual H\ped{x}TiSe\ped{2} samples are likely to be characterized by a strong disorder in the distribution of H atoms, and will thus exhibit a mixture of different single phases which however all turn out to be dynamically unstable (see Supplementary Note 8 for further details). This in turn prevents a full \emph{ab-initio} calculation of the properties of the real system, in particular for the calculation of the SC critical temperature. 

To have solid first-principles predictions of the SC phase even in the high density regime, the metallic $1T$-H\ped{1}TiSe\ped{2} phase was therefore artificially stabilized by slightly reducing the lattice constants by 4\%  (that corresponds to an external pressure of $\sim$\,10\,GPa, see Methods). This approach, already exploited to remove the CDW distortion in pure $1T$-TiSe\ped{2}\,\cite{PhysRevLett.106.196406} and reduce anharmonic effects in H-doped palladium alloys\,\cite{VocaturoJAP2022}, allows obtaining a dynamically stable high-doping phase while at the same time leaving the main features of the electronic properties of $1T$-H\ped{1}TiSe\ped{2} mostly unaffected (as shown in Supplementary Fig.\,11).
The resulting phonon band structure shows the appearance of H-derived branches (highlighted in red in the phonon density of states shown in Fig.\,\ref{figure:theory}d), comprising a high-energy band around 1300\,cm\apex{-1} and more entangled branches at lower frequencies, close to the TiSe\ped{2} modes around 500\,cm\apex{-1}. Overall, the pressurized H\ped{1}TiSe\ped{2} phase is therefore able to mimic all the key modifications introduced by the H dopants in the real TiSe\ped{2} samples.

In particular, the dynamical stability of the pressurized H\ped{1}TiSe\ped{2} structure allows computing the electron-phonon spectral function $\alpha^2F(\omega)$, shown in Fig.\,\ref{figure:theory}d, from which both the electron-phonon coupling $\lambda$ and the SC $T\ped{c}$ can be determined (Methods). In sharp contrast with the high-pressure hydrides\,\cite{FloresLivasPR2020, BoeriJPCM2021}, the H-derived phonon branches do not sensibly contribute to the total coupling $\lambda=1.15$, which is however found to be significantly higher with respect to that of the low-doping structure ($\lambda=0.5$, see Supplementary Note 8). As a consequence, H\ped{1}TiSe\ped{2} is predicted to become superconducting at $T\ped{c}\sim13$\,K (Methods), a critical temperature surely higher than the experimentally-measured one, but of the same order of those measured in other SC TMDs\,\cite{LiNatCommun2021, PhysRevB.92.134510, PhysRevLett.119.087003}. This overestimation can be ascribed to differences between the real material and the computational model, with disorder effects and doping inhomogeneities easily lowering the real $T\ped{c}$ with respect to the ideal case considered in the calculations. Anyway, the much larger predicted $T\ped{c}$ with respect to that of the low-doping structure ($T\ped{c}\sim 1.5$\,K, see Supplementary Note 7) highlights the crucial role played by the H dopants in determining the properties of the SC phase. Note that this enhancement cannot be ascribed merely to the application of pressure to the TiSe\ped{2} structure, since the $T\ped{c}$ of undoped $1T$-TiSe\ped{2} at 10\,GPa is $\ll1$\,K\,\cite{KusmartsevaPRL2009, PhysRevLett.106.196406}. 

\section*{Discussion}

Our results show that ionic liquid gating-induced protonation allows for a robust and non-volatile tuning of the electronic ground state of archetypal correlated layered compound $1T$-TiSe\ped{2}. We demonstrate that this tuning is qualitatively different from what is obtained by other doping methods, such as electrostatic gating and Cu or Li intercalation. An increased level of hydrogen doping suppresses the intensity of the charge-density wave phase of undoped TiSe\ped{2} and triggers the onset of superconductivity. 
Still, vestigial traces of the charge-density wave ordering, observed even in H\ped{2}TiSe\ped{2} -- a stoichiometry unattainable via Li or Cu intercalation -- hint at the coexistence of both phases in a wide doping range{\color{blue}. This coexistence is likely enabled by the strongly disordered nature of the hydrogen-rich compound, which hosts a mixture of different phases with distinct electronic structures determined by the several possible hydrogen intercalation configurations.}
The observed superconducting phase, possibly gapless and multi-band in character, together with the absence of an anomalous metallic phase are also at odds with what is reported in other superconducting TiSe\ped{2} compounds. Our experimental results are consistent with \textit{ab initio} calculations showing how the effect of H intercalation is not limited to a rigid-band doping, but can lead to a full band-structure engineering of the undoped TiSe\ped{2} compound.
The role of hydrogen is crucial in determining the emergence of superconductivity through different mechanisms: i) at low concentrations, H intercalation rigidly charge-dopes the system, weakening the CDW ordering and increasing the electron-phonon coupling at the Fermi level; ii) at higher concentrations, the doping mechanism changes, triggering a band inversion that cannot be described within a rigid band shift; this leads to the partial filling of the highly-coupled $d_{z^2}$ band, mimicking the exact band population found in Nb-based transition-metal dichalcogenides of the $2H$ polytype, such as $2H$-NbSe$_2$ and $2H$-NbS$_2$, which are superconductors even without doping; iii) H intercalation also induces the hybridization of these $d_{z^2}$ states with H-derived orbitals, reducing the screening of the phonon perturbations and thereby further increasing the electron-phonon coupling.
This clearly indicates that the superconducting phase realized in H\ped{x}TiSe$_2$ should be of a different nature with respect to those observed in Cu\ped{x}TiSe$_2$ or in electron-doped MoS$_2$, reflecting the different orbitals involved in the electron-electron pairing and the capability of H doping to hybridize the electronic states at the Fermi level and reduce the screening to phonon perturbations. 
At the same time, its origin is firmly distinct from that found in high-pressure hydrides, since in H\ped{x}TiSe$_2$ the high-frequency phonon spectral contributions due to the light H atoms remain weakly coupled to the electronic states at the Fermi level.

In summary, our findings show ionic liquid gating-induced protonation to be a unique doping technique, distinct from the more ubiquitous chemical substitution and alkali intercalation. 
Furthermore, the understanding of the difference between the mechanisms at play in hydrogen-rich superconductors at megabar pressures\,\cite{FloresLivasPR2020} and in H\ped{x}TiSe\ped{2} will be crucial in the quest for new high-T\ped{c} superconductors at ambient and low pressures, as also suggested by the room-temperature superconductivity recently claimed to appear in lutetium hydride at relatively low pressures\,\cite{Dias2023}. 
Gate-driven protonation will enable accessing new electronic phases of the host compounds at ambient pressure, where hydrogen not only injects high-frequency phonon spectral contributions, but can play a triple role as a charge dopant, a source of band-selective filling, and a knob to tune the electronic screening via orbital hybridization. 

Theoretical proposals in hydrogen doped systems are recently appearing, confirming our experimental and theoretical analysis. All studies highlight how superconductivity can be induced in different materials by hydrogen incorporation. Among others, we cite the hydrogen-induced superconducting phases in PdCu\,\cite{VocaturoJAP2022}, diborides like MgB\ped{2}\,\cite{BekaertPRL2019} and TiB\ped{2}\,\cite{WangIJMPC2021}, chromium\,\cite{YuSciRep2015}, Mo\ped{2}C\ped{3}\,\cite{JiaoEPL2022}, hBN\,\cite{RawalPRM2022}, and in carbon nanostructures\,\cite{SannaEPJB2018}.
The experimental realization of these theoretical proposals, made possible by the simple and reliable technique shown in this work, can reverse the accepted paradigm to access high-temperature superconducting phases of hydrides from the high-pressure side. Here we propose an alternative route for stabilizing new hydrogen-rich compounds, starting from promising materials at ambient pressure, driving them to the predicted superconducting state and then, eventually, further enhancing their superconducting properties by applying low or moderate pressures.
	
\section*{Methods}

\begin{footnotesize}

\textbf{Ionic liquid gating-induced protonation.}
Freshly-cleaved 1$T$-TiSe\ped{2} crystals (HQ Graphene\tcb{; typical size 1.0$\times$0.5$\times$0.05\,mm\apex{3}}) were electrically contacted by drop-casting small droplets of silver paste (RS Components) to thin gold wires and immersed in a Duran crucible \tcb{(40\,mm diameter)} filled with 1-ethyl-3-methylimidazolium tetrafluoroborate ionic liquid (EMIM-BF\ped{4}, Sigma Aldrich) together with a platinum (Pt) counter electrode. The gate voltage $V\ped{G}$ was applied to the Pt electrode at 300\,K in ambient atmosphere by an Agilent B2961 power source. {\color{blue}The applied $V\ped{G}$ was always limited to a maximum value of +3\,V and the gating temperature to 300\,K so as to avoid undesired electrochemical reactions between the sample and the ionic liquid such as sample etching\,\cite{ShiogaiNatPhys2016} or organic-ion intercalation\,\cite{PiattiNanomaterials2022}.} The four-wire resistance $R=V\ped{xx}/I\ped{DS}$ was monitored in situ by sourcing a constant current $I\ped{DS}\approx 100\,\upmu$A between the outer drain (D) and source (S) contacts with a Keithley 220 current source, and measuring the longitudinal voltage drop $V\ped{xx}$ between the inner voltage contacts with an HP3457 multimeter. The resistivity was then determined as $\rho = R t w l^{-1}$, where $t$ and $w$ are the sample thickness and width, and $l$ is the distance between the inner voltage contacts. Common-mode offsets were removed using the current-reversal method. The protonated TiSe\ped{2} samples were then extracted from the cell and rinsed with acetone and ethanol to remove ionic-liquid residues before further \textit{ex-situ} characterizations. In between measurements, samples were stored in standard desiccators either under low vacuum or in argon atmosphere to avoid moisture contamination.\\

\textbf{Electric transport measurements.}
All protonated TiSe\ped{2} crystals discussed in this work were characterized via \tcb{\textit{ex-situ}} temperature-dependent resistivity measurements carried out in the high-vacuum chamber of a Cryomech pulse-tube cryocooler with a base temperature of $\approx 2.8\,$K. The magnetotransport properties of selected crystals were measured either up to $\approx 4.5\,$K in the variable-temperature insert of a \apex{4}He Oxford cryostat, or up to $\approx 300\,$K in a Quantum Design physical properties measurement system (PPMS), both equipped with 9\,T superconducting magnets. The resistivity was determined as in the room-temperature gating runs, except that the source-drain current was sourced via an Agilent B2912 source-measure unit (Oxford cryostat) or using a standard 4-probe technique with the Resistivity option for the PPMS. The user bridge board of the PPMS has a digital-to-analog converter (DAC) which adjusts the excitation current and a delta-sigma A/D converter which reads the voltage output. In AC mode, the user bridge board applies a DC excitation to the sample and at each measurement it reverses the current averaging the absolute value of the positive and negative voltage readings. This operation eliminates errors from DC offset voltages and produces the most accurate readings.
Both the longitudinal ($V\ped{xx}$) and transverse ($V\ped{xy}$) voltage drops were measured via an Agilent 34420 nanovoltmeter (Oxford cryostat) or by means of a standard 6-terminal method with the Resistivity option for the PPMS. The magnetic field $H$ was always applied orthogonal to the samples' \textit{ab} plane. 

The Hall coefficient was determined from the antisymmetrized $R\ped{xy} = V\ped{xy}/I\ped{DS}$ data as:
\begin{equation}
R\ped{H}=\frac{t}{\mu_0}\times\frac{dR\ped{xy}(T)}{dH}
\end{equation}
and the Hall density was calculated as $n\ped{H}=(eR\ped{H})^{-1}$ where $e$ is the elementary charge.

In superconducting samples, \tcb{the residual resistivity ratio was determined as $\rho(300\,\mathrm{K})/\rho_0$, where $\rho_0$ is the resistivity value in the normal state immediately above the superconducting transition;} the onset transition temperature $T\ped{c}\apex{on}$ was determined as the threshold where $\rho(T)$ reaches 95\% of $\rho_0$; and the zero-temperature in-plane coherence length was determined as: 
\begin{equation}
	\xi\ped{ab}(0) = \sqrt\frac{\Phi_0/2\pi}{-\mu_0(dH\ped{c2}/dT)\, T\ped{c}\apex{on}}
\end{equation}
where $H\ped{c2}$ is the magnetic field at which $\rho(H)$ reaches 95\% of its normal-state value, $\mu_0$ is the vacuum permeability, $dH\ped{c2}/dT$ is the slope of the $H\ped{c2}{-}T$ curve, and $\Phi_0$ is the magnetic flux quantum.\\

\textbf{X-ray diffraction.}
X-ray powder diffraction patterns of the TiSe$_2$ samples were collected \tcb{\textit{ex situ} on selected single crystals after thorough removal of the ionic liquid residues and of the electrical leads} using a $114.6\,{\rm mm}$ Gandolfi camera, with Ni-filtered Cu K-$\alpha$ radiation source and an exposure time of 48 hours. Diffraction was impressed on a photographic film and reduced using the software X-RAY\,\cite{xray} to obtain the intensity profiles as a function of the diffraction angle.\\

\textbf{Raman spectroscopy.}
Raman spectra were acquired on freshly-cleaved surfaces in ambient conditions using a Renishaw InVia H43662 micro-Raman spectrometer. All spectra were acquired using an excitation wavelength of 514\,nm, a laser power $<$\,1\,mW focused through a 100X objective, an exposure time of 20\,s, and 50 accumulations.\\

\textbf{Nuclear magnetic resonance.}
{}$^{1}$H-NMR measurements at $\mu_{0}H \simeq 3.5$\,T (magnetic field parallel to the crystallographic $ab$ plane) were performed using a TecMag Apollo spectrometer coupled to a resonant circuit made of an in-series combination of a $470$ pF capacitor and of a seven-loop solenoidal coil. The coil was deformed in order to mimic the flake-like shape of the crystals and to maximize the geometrical filling factor, in turn. The spin-lattice relaxation time T$_{1}$ was quantified via a conventional inversion-recovery pulsed sequence. In particular, \Tone\ was extracted based on a best-fitting of the experimental recovery curves based on a stretched-exponential function
\begin{equation}
	M(\tau) = M(\infty) \left\{1-2f\exp\left[-\left(\frac{\tau}{T_{1}}\right)^{\beta}\right]\right\}.
\end{equation}
Here, $M(\tau)$ is the component of the nuclear magnetization along the quantization axis at time $\tau$ after the first inversion radio-frequency pulse, $M(\infty)$ is the equilibrium magnetization, $f \leq 1$ allows for non-ideal inversion conditions and $\beta$ is the stretching exponent. $\beta \sim 0.9$ was quantified independently on temperature, in good agreement with the purely-exponential behaviour expected for spin-$1/2$ nuclei.\\
	
\textbf{Magnetization measurements.}
Systematic measurements of the magnetic moment as a function of temperature and magnetic field were carried out on a small H\ped{2}TiSe\ped{2} single crystal \tcb{with approximate surface $\sim 1$\,mm$^{2}$} by means of a superconducting quantum interference device (SQUID) magnetometer. Since the sample mass was less than 0.1\,mg (i.e., below the sensibility threshold of a standard laboratory balance), the magnetic field had to be applied perpendicular to the sample surface (i.e., to $ab$ planes). Such setup maximizes the effects of the demagnetizing factor, thus enhancing the diamagnetic signal in the superconducting state.
The unavailability of a reliable value for the sample mass (or volume) prevents us from providing an accurate estimate of the shielding fraction in the superconducting state.\\

\textbf{Muon-spin rotation measurements.}
Muon-spin rotation (\musr) is an extremely sensitive probe of the local (i.e., microscopic) electronic properties, which uses spin-polarized positive muons implanted in the sample under test\,\cite{Blundell1999, Amato1997}. Depending on the material density, muons typically penetrate over a depth of several hundreds of microns and, thus, are implanted homogeneously over the whole sample volume. Because of this, muons are  
considered as a bulk probe of matter. 
All the \musr\ measurements were carried out at the Dolly spectrometer ($\pi$E1 beamline) of the Swiss Muon Source at the Paul Scherrer Institute, Villigen, Switzerland. A $^{3}$He cryostat was used to reach temperatures down to 0.28\,K. A mosaic of H\ped{2}TiSe\ped{2} single crystals was glued using GE-varnish on a 25-$\upmu$m thick copper foil, here acting as a thermal link \textcolor{blue}{(Supplementary Fig.\,7)}. 
A 100-$\upmu$m-thick high-purity silver degrader was fixed in front of the sample. All measurements were performed using an active veto scheme, thus removing the background signal due to muons which miss the sample. 
\musr\ measurements were performed following both standard field-cooling (FC) and nonstandard zero-field cooling (ZFC) protocols. In the former case, a regular lattice of vortex lines is established over the sample volume. This is a key 
requirement in order to accurately determine the magnetic penetration depth and the superfluid density. 
In the ZFC case, instead, we induce an artificially disordered vortex lattice in the SC state that, in turn, causes highly inhomogeneous local magnetic fields at the implanted muon sites and, hence, a much higher muon-spin depolarization rate than that observed in a conventional FC- (or normal-state) experiment. This procedure is particularly suited for evaluating the superconducting volume fraction of superconductors with a weak magnetic-flux expulsion, as the H-doped TiSe$_2$ in the present case.\\

\textbf{Muon spin rotation fits.}
In all cases, the time-dependent TF polarization was fitted using the following function:
 \begin{equation}
 	\begin{split}\label{eq:pt}
 		P_\mathrm{TF}(t) &=f_\mathrm{sc}\cos(\gamma_{\mu}B_{\mu}t+\phi)\exp\left(-\sigma^{2}t^{2}/2\right)\\
 		&+f_\mathrm{tail}\cos(\gamma_{\mu}B_\mathrm{tail}t+\phi)\exp(-\Lambda t)
 	\end{split}
 \end{equation}

Here, $f_\mathrm{tail} = 1 - f\ped{sc}$ is the fraction of muons implanted in the silver 
degrader and in the non superconducting parts of the sample, $\gamma_{\mu}/2\pi= 135.53$\,MHz/T is the muon gyromagnetic ratio, $B_{\mu}$ and $B_\mathrm{tail}$ are the magnetic fields probed by the muons implanted in the sample and in the silver degrader, and $\phi$ is a common initial phase. Finally, $\Lambda$ is the relaxation rate originating from the muons implanted in the silver degrader.

Both $\Lambda$ and $f_\mathrm{tail}$ parameters were obtained in a two-step calibration via a 
ZFC TF experiment at 0.28\,K (see Fig.\,\ref{figure:4}). First, the long-time tail ($t>5$\,$\upmu$s) 
was fitted with a single oscillating component ($f_\mathrm{sc}=0$), determining $\Lambda=30\pm6$\,ns$^{-1}$. In this case, due to the fast signal decay in the SC phase, the long-time tail reflects only those muons implanted in the silver degrader (and in non-superconducting parts of the sample). Secondly, by fixing $\Lambda$ to the previously determined value, the data were fitted over the whole time range to obtain $f_\mathrm{tail}=0.56\pm0.02$. These two parameters were then kept fixed during all the subsequent TF-fits.

From the SC depolarization rate $\sigma\ped{sc}$, under the assumption of an ideal triangular vortex lattice, and by considering that the applied field is negligibly small with respect to the upper critical field determined from resistivity and magnetization measurements ($H_{c2}>0.1$\,T), the magnetic penetration depth is determined as $\lambda= (0.00371\,\Phi_{0} \gamma_{\mu}^{2}/ \sigma_\mathrm{sc}^{2})^{1/4}$ 
\cite{Barford1988,Brandt2003}.\\

\textbf{Density functional theory calculations.}
First-principles calculations were performed using the \textsc{Quantum Espresso} package\,\cite{QEcode, QE-2017}. 
For all calculations, the experimental lattice parameters and the Generalized Gradient Approximation (GGA) for the exchange and correlation energy were adopted. This is because in 1$T$-TiSe$_2$ the GGA functional, on top of the experimentally determined lattice constants, successfully predicts both the internal structural parameters (Se-Ti distance) and the dynamical instabilities guiding the system to the $2\times2\times2$ charge density wave phase\,\cite{PhysRevB.92.094107}, in perfect agreement with experiments. 
Conversely, since the GGA functional does not describe perfectly the electronic properties of TiSe\ped{2} due to the strong correlation effects arising from the localized $d$ orbitals of Ti\,\cite{Rohwer2011, PhysRevB.92.094107}, the effects of local correlations were accounted for by including a Hubbard-like correction for Ti-$d$ orbitals in a GGA+$U$ approach\,\cite{PhysRevB.44.943}. The $U$ parameter, determined from first principles\,\cite{PhysRevB.71.035105, PhysRevB.92.094107}, was set to $U=3.9$\,eV, allowing to reproduce the Fermi surface \tcb{experimentally determined via angle-resolved photoemission spectroscopy\,\cite{Rohwer2011, RaschPRL2008, WatsonPRL2019}}.
Electron-ion interaction was described with pseudopotentials which include the 3$s$, 3$p$, 4$s$ and 3$d$ valence states for Ti and 4$s$, 4$p$ and 4$d$ for Se, with an energy cutoff up to 70\,Ry (840\,Ry for charge density).
The integration over the Brillouin Zone (BZ) was performed using a uniform 24$\times$24$\times$14 grid (for the TiSe$_2$ unit cell, and properly rescaled for the supercell calculations) with a 0.005\,Ry Gaussian smearing. The electronic density of states are evaluated sampling the BZ on a 30$\times$30$\times$18 uniform grid using tetrahedron method for integration\,\cite{PhysRevB.89.094515}.
Several structural models, starting from low concentration regime up to higher doping, were considered using a random-search algorithm (over 100 configurations) to determine the H atom's lowest-energy intercalation site.
One H atom is placed in a 2$\times$2$\times$2 1$T$-TiSe$_2$ supercell in the lowest concentration ($\mathrm{x}=0.125$), whereas two H atoms are included in the 1$\times$1$\times$1 unit cell ($\mathrm{x}=2$) in the highest. 
In the low concentration regime, we unfold the electronic bands onto the 1$T$-TiSe$_2$ BZ by means of the Bands-UP software\,\cite{PhysRevB.89.041407,PhysRevB.91.041116}.
The application of an external pressure of $\sim 10$\,GPa was realized by reducing the lattice constants of about 4\% at constant $c/a$ ratio, where $c$ and $a$ are the out-of-plane and in-plane lattice constants respectively.
Phonon calculations were performed using the Density Functional Perturbation Theory (DFPT)\,\cite{PhysRevA.52.1086} in the harmonic approximation sampling the reciprocal space with a $4\times4\times4$ grid. The electron-phonon coupling was converged using a $48\times48\times28$ k-point grid for the electronic momenta. The superconducting critical temperature was estimated using the Allen-Dynes modified Mc-Millan equation\,\cite{Dynes1972,PhysRevB.12.905} with $\mu^*=0.1$\,\cite{PhysRevLett.106.196406,PhysRevB.96.165404} (see also Supplementary Note 8).

\section*{Data availability}
The data that support the findings of this study are available from the authors upon reasonable request.

\end{footnotesize}

\bibliographystyle{naturemag}
\bibliography{bibliography}

\bigskip
\bigskip

\begin{footnotesize}

\section*{Acknowledgments}
E.P., S.R., G.Profeta, D.D. and R.S.G. acknowledge support from the MIUR PRIN-2017 program (Grant No.2017Z8TS5B -- “Tuning and understanding Quantum phases in 2D materials -- Quantum2D”). We thank S. Guastella and M. Bartoli for their assitance in the XPS measurements and analysis. G. Prando acknowledges useful discussions with M. Filibian and D. Ravelli concerning the {}$^{1}$H-NMR proton quantification procedure. S.R. is grateful to the Center for Instrument Sharing and the Department of Earth Sciences of the University of Pisa for the support in the XRD measurements and analysis.


\section*{Author contributions}
E.P., G.Profeta and R.S.G. conceived the idea.
R.S.G. directed the project.
E.P., D.D., and R.S.G. designed and performed the protonation and the electric transport measurements.
G.Prando and P.C. performed the NMR characterization.
M.M. and M.P. contributed to the electric transport measurements.
C.T. and G.Profeta performed the \emph{ab initio} calculations.
S.R. provided the pristine samples and performed the XRD characterization.
G.L. performed the magnetization measurements.
G.Prando, G.L., and T.S. performed the \musr\ measurements.
All authors contributed to the discussion and interpretation of the results.
E.P., G.Prando, G.L., T.S., G.Profeta, D.D., and R.S.G. wrote the manuscript with input from all authors.\\
		
\section*{Competing interests}
The authors declare no competing interests.
\end{footnotesize}
 
\end{document}


\title{Supplementary Information for\texorpdfstring{\\}{ } 
Superconductivity induced by gate-driven hydrogen intercalation\texorpdfstring{\\}{ } 
in the charge-density-wave compound \texorpdfstring{$1T$-TiSe$_2$}{1T-TiSe2}
}

\author{Erik Piatti}
\email{erik.piatti@polito.it}
\affiliation{Department of Applied Science and Technology, Politecnico di Torino, I-10129 Torino, Italy}
\author{Giacomo Prando}
\affiliation{Department of Physics, Università di Pavia, I-27100 Pavia, Italy}
\author{Martina Meinero}
\affiliation{Consiglio Nazionale delle Ricerche-SPIN, I-16152 Genova, Italy}
\affiliation{Department of Physics, Università di Genova, I-16146 Genova, Italy}
\author{Cesare Tresca}
\affiliation{Department of Physical and Chemical Sciences, Università degli Studi dell’Aquila, I-67100 L’Aquila, Italy}
\affiliation{SPIN-CNR, Università degli Studi dell'Aquila, I-67100 L'Aquila, Italy}
\author{Marina Putti}
\affiliation{Consiglio Nazionale delle Ricerche-SPIN, I-16152 Genova, Italy}
\affiliation{Department of Physics, Università di Genova, I-16146 Genova, Italy}
\author{Stefano Roddaro}
\affiliation{Istituto Nanoscienze-CNR, NEST and Scuola Normale Superiore, I-56127 Pisa, Italy}
\author{Gianrico~Lamura}
\affiliation{Consiglio Nazionale delle Ricerche-SPIN, I-16152 Genova, Italy}
\author{Toni Shiroka}
\affiliation{Laboratory for Muon-Spin Spectroscopy, Paul Scherrer Institut, CH-5232 Villigen PSI, Switzerland}
\affiliation{Laboratorium f{\"u}r Festk{\"o}rperphysik, ETH Z{\"u}rich, CH-8093 Zurich, Switzerland}
\author{Pietro Carretta}
\affiliation{Department of Physics, Università di Pavia, I-27100 Pavia, Italy}
\author{Gianni Profeta}
\affiliation{Department of Physical and Chemical Sciences, Università de L’Aquila, I-67100 L’Aquila, Italy}
\affiliation{SPIN-CNR, Università de L'Aquila, I-67100 L'Aquila, Italy}
\author{Dario Daghero}
\affiliation{Department of Applied Science and Technology, Politecnico di Torino, I-10129 Torino, Italy}
\author{Renato S. Gonnelli}
\email{renato.gonnelli@polito.it}
\affiliation{Department of Applied Science and Technology, Politecnico di Torino, I-10129 Torino, Italy}

\maketitle


\clearpage

\widetext

\section*{Supplementary Note 1: X-ray photoelectron spectroscopy of the titanium diselenide crystals}

We employed X-ray photoelectron spectroscopy (XPS) to confirm 
that the ionic liquid gating induced protonation did not result in the intercalation of undesired species in the TiSe\ped{2} lattice, such as the cations of the ionic liquid itself\,\cite{WangCPB2021, PiattiNanomaterials2022}.
XPS spectra were acquired with a PHI Model 5000 electron spectrometer equipped with an aluminum anode (1486\,eV) monochromatic source, with a power of 25\,W
, at pressures below $5\times10^{-8}$\,mbar. All crystals were cleaved immediately before loading them in the XPS load-lock, preliminarily degassed overnight at room temperature, and a surface cleaning step was performed by etching the first few atomic layers via sputtering with the built-in Ar-ion gun. The binding energies of the acquired spectra were corrected for specimen charging by referencing the C\,1$s$ line (acquired in dedicated survey scans before the Ar cleaning procedure) to 284.8\,eV. Peak assignments were obtained using the MultiPak Software 9.7 for peak recognition.

\begin{figure}[h] 
	\centering
	\includegraphics[width=0.7\textwidth]{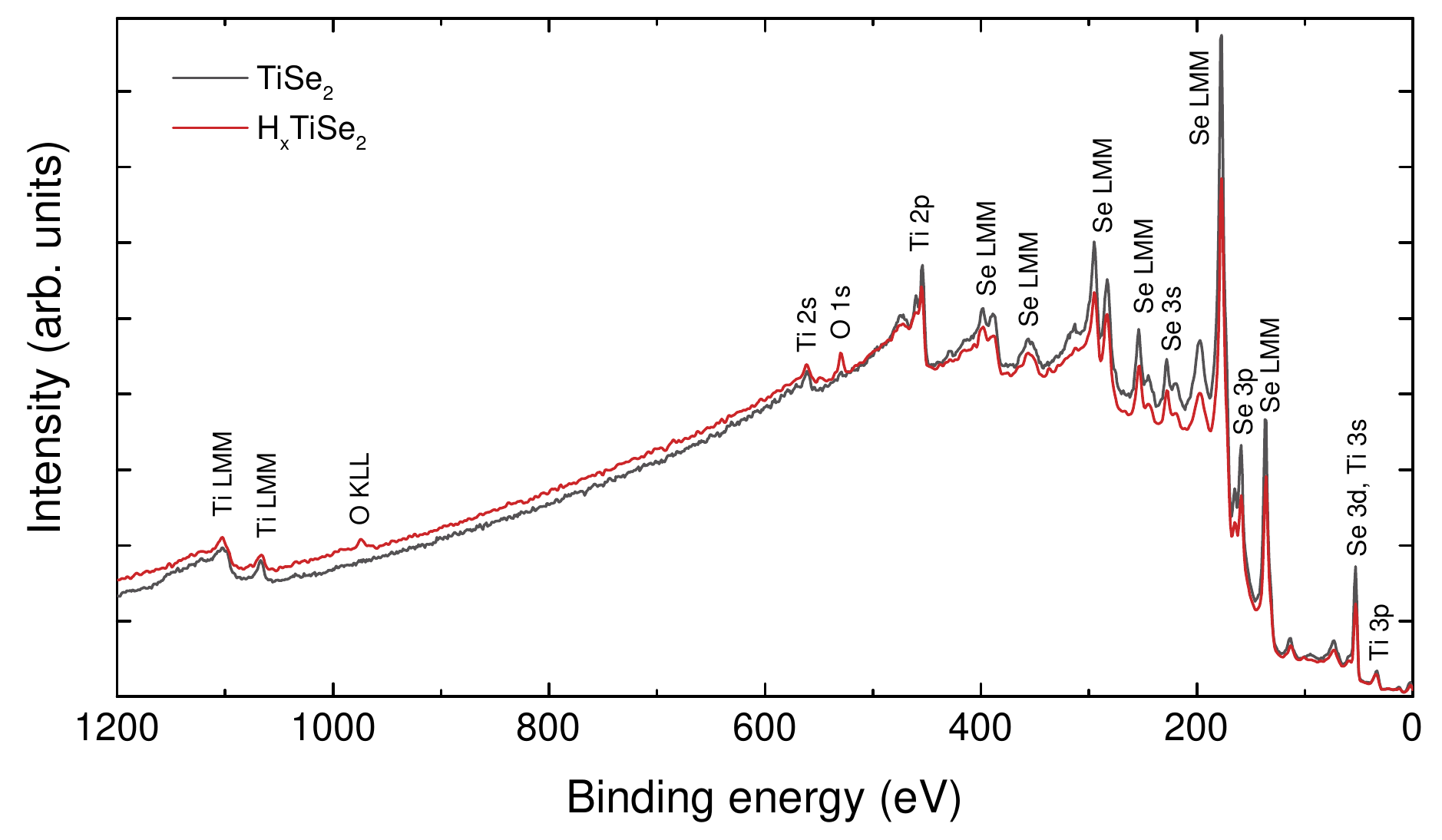}
	\caption{\label{fig:XPS}
		X-ray photoelectron spectra of the TiSe\ped{2} crystals after being Ar-cleaned in high vacuum, together with the peak assigments obtained using the MultiPak Software 9.7 for peak recognition. The solid black line is the survey scan up to 1200\,eV of the pristine TiSe\ped{2} crystal, the solid red line is that of the H\ped{2}TiSe\ped{2} crystal.}
\end{figure}

Supplementary Fig.\,\ref{fig:XPS} shows the full survey scans up to 1200\,eV acquired on a pristine TiSe\ped{2} crystal (solid black line) and a H\ped{2}TiSe\ped{2} crystal (solid red line) after the surface cleaning by Ar-ion sputtering, together with the corresponding peak assigments. Only peaks arising from Ti, Se, and O atomic species are observed in both scans. The Ti and Se signals obviously arise from the TiSe\ped{2} lattice, whereas the presence of an O signal indicates a small degree of oxidation even in the pristine crystal, which is nearly unavoidable when any TiSe\ped{2} surface is exposed to ambient air\,\cite{SunACIE2017}. The intensity of the O signal is found to increase in the H\ped{2}TiSe\ped{2} crystal: we ascribe this to an enhanced level of oxidation in the exposed TiSe\ped{2} surfaces triggered by the harsh electrochemical environment of the protonation cell, which cannot be fully removed by the Ar cleaning treatment across the whole area probed by the X-ray beam.
Most importantly, no additional peaks ascribable to the atomic species building up the EMIM-BF\ped{4} ionic liquid molecular structure (C, N, B, or F) are detected in the H\ped{2}TiSe\ped{2} crystal, thereby confirming a negligible incorporation (if any) of either the EMIM\apex{+} cation or the BF\ped{4}\apex{-} anion in the TiSe\ped{2} lattice structure, in full agreement with the minute unit cell expansion detected by X-ray diffraction as discussed in the Main Text.

\clearpage

\section*{Supplementary Note 2: Quantification of intercalated protons in the titanium diselenide crystals}\label{SectQuantification}

We quantified the number of protons in the investigated collection of TiSe$_{2}$H$_{x}$ crystals, with overall mass $(2.90 \pm 0.06)$ mg and volume $\sim 3 \times 3 \times 0.5$ mm$^{3}$, by means of {}$^{1}$H nuclear magnetic resonance (NMR). We prepared a resonant circuit using an in-series combination of a $470$ pF capacitor and of a seven-loop solenoidal coil which was deformed in order to mimic the flake-like shape of the crystals and to maximize the geometrical filling factor in turn. We also prepared a flake-like reference sample of hexamethylbenzene (C$_{12}$H$_{18}$), with similar mass $(3.16 \pm 0.05)$ mg and dimensions if compared to the TiSe$_{2}$H$_{x}$ crystals, aiming at a comparable geometrical filling factor of the solenoidal coil.

\begin{figure}[h]
	\centering
\begin{tikzpicture}[
	mybox/.style={
		draw=black,
		fill=black!20,
		minimum size=1cm
	},
	axis/.style={
		thick,
		-stealth
	},
	annotation/.style={
		latex-latex
	},
	myplot/.style={
		domain=0:T/2,
		smooth,
		line width = 1pt
	},
	declare function={
		T=16;
		RFheight=2.5;
		Aqheight=0;
		plotshift=0.5;
		Tmax=16.5;
		Tmin=-1;
	}
	]
	
	\draw (14.2cm,Aqheight) -- (-1cm, Aqheight) node[left] {Aquisition};
	\draw (14.2cm,RFheight) -- (-1cm, RFheight) node[left] {RF Pulse};
	
	\node [mybox, minimum width=1cm] (b) at (plotshift,RFheight) {};
	\node [mybox, minimum width=1cm] (c) at (plotshift+3cm,RFheight) {};
	\node [mybox, minimum width=5cm] (d) at (plotshift+8cm,Aqheight) {};
	\node (e) at (plotshift+14.2cm,Aqheight-0.5cm) {};
	
	\draw [dashed] (b.south west) -- ++(0,-0.5) coordinate(tmpa);
	\draw [dashed] (b.south east) -- ++(0,-0.5) coordinate(tmpb);
	\draw [annotation] (tmpa) -- node[below] {$\tau\ped{pulse}$} (tmpb);
	
	\draw [dashed] (b.north) -- ++(0,0.5) coordinate(tmpa);
	\draw [dashed] (c.north) -- ++(0,0.5) coordinate(tmpb);
	\draw [annotation] (tmpa) -- node[above] {$\tau\ped{e}$} (tmpb);
	
	\draw [dashed] (c.south west) -- ++(0,-0.5) coordinate(tmpa);
	\draw [dashed] (c.south east) -- ++(0,-0.5) coordinate(tmpb);
	\draw [annotation] (tmpa) -- node[below] {$\tau\ped{pulse}$} (tmpb);
	
	\draw [dashed] (c.north) -- ++(0,0.5) coordinate(tmpa);
	\draw [dashed] (d.north west) -- ++(0,3) coordinate(tmpb);
	\draw [annotation] (tmpa) -- node[above] {$\tau\ped{e}$} (tmpb);
	
	\draw [dashed] (d.north west) -- ++(0,3) coordinate(tmpa);
	\draw [dashed] (d.north east) -- ++(0,3) coordinate(tmpb);
	\draw [annotation] (tmpa) -- node[above] {$\tau\ped{a}$} (tmpb);
	
	\draw [dashed] (d.north east) -- ++(0,3) coordinate(tmpa);
	\draw [thick,double] (e.south) -- ++(0,4.125) coordinate(tmpb);
	\draw [annotation] (tmpa) -- node[above] {$\tau\ped{r}$} (tmpb);
	
\end{tikzpicture}
	\caption{\label{FigNMRdiag} Diagram of the two-pulse solid-echo sequence employed in the {}$^{1}$H nuclear magnetic resonance measurements. Pulse duration $\tau\ped{pulse}$, separation time $\tau\ped{e}$, acquisition time $\tau\ped{a}$, and idle time $\tau\ped{r}$ are highlighted.}
\end{figure}
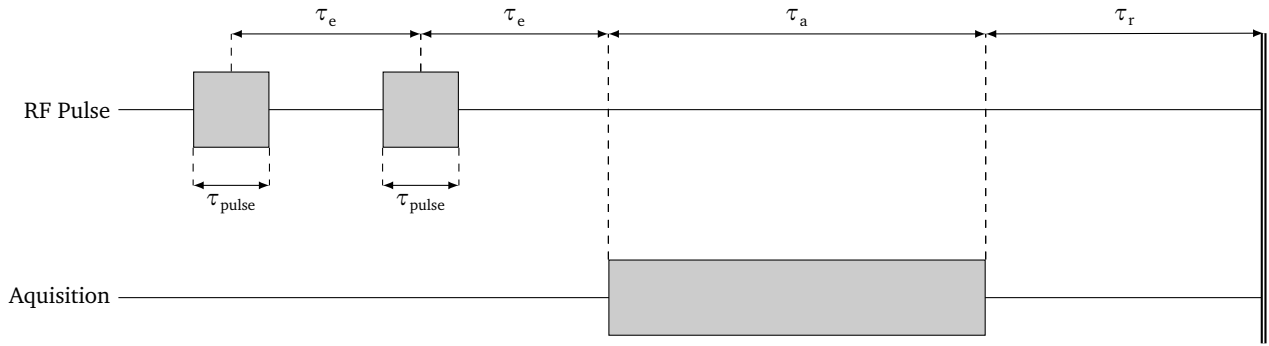

The proton quantification was performed at constant temperature ($T = 292.5$ K) and magnetic field $\mu_{0}H \simeq 3.46$ T, corresponding to a Larmor frequency $\nu\ped{L} \simeq 147.83$ MHz for the {}$^{1}$H nuclear magnetic moments. The procedure relied on the two-pulse solid-echo sequence shown in Supplementary Fig.\,\ref{FigNMRdiag}. The first radiofrequency (RF) pulse generates a free-induction decay signal which is refocused into a spin-echo signal by the second pulse. Both pulses have the same duration $\tau\ped{pulse}$. The spin-echo is centred at a time $2\tau\ped{e}$ after the first pulse, $\tau\ped{e}$ being the time separation between the two pulses. After the acquisition time $\tau\ped{a}$, the idle time $\tau\ped{r}$ is such that $\tau\ped{r} \gtrsim 5 T_{1}$, where $T_{1}$ is the spin-lattice relaxation time of the considered sample. This latter calibration guarantees that the system reaches conditions of thermodynamical equilibrium before the sequence is started over again and repeated until a satisfactory signal/noise ratio is reached.

\begin{figure}[b!] 
	\centering
	\includegraphics[width=0.48\textwidth]{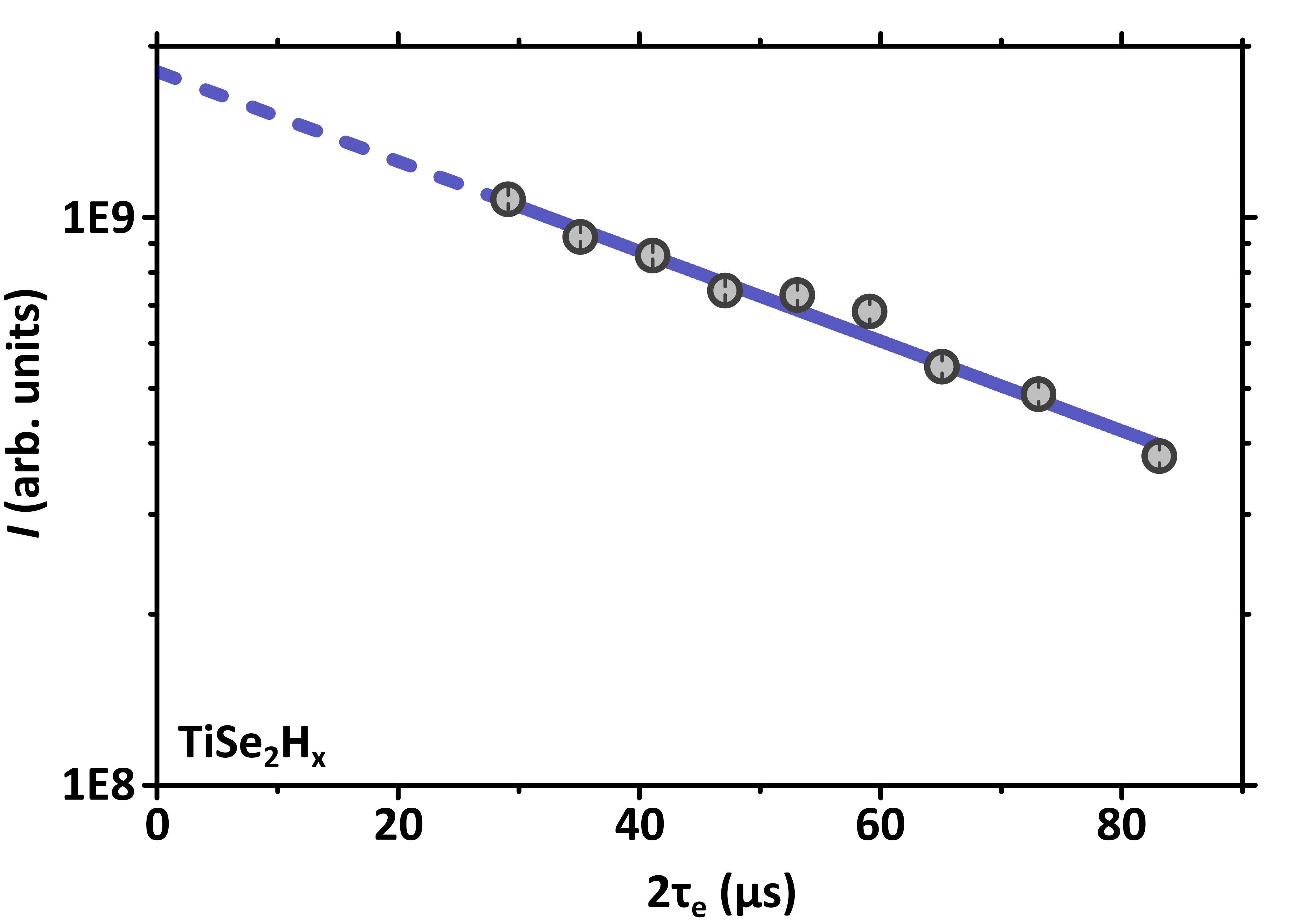} \includegraphics[width=0.48\textwidth]{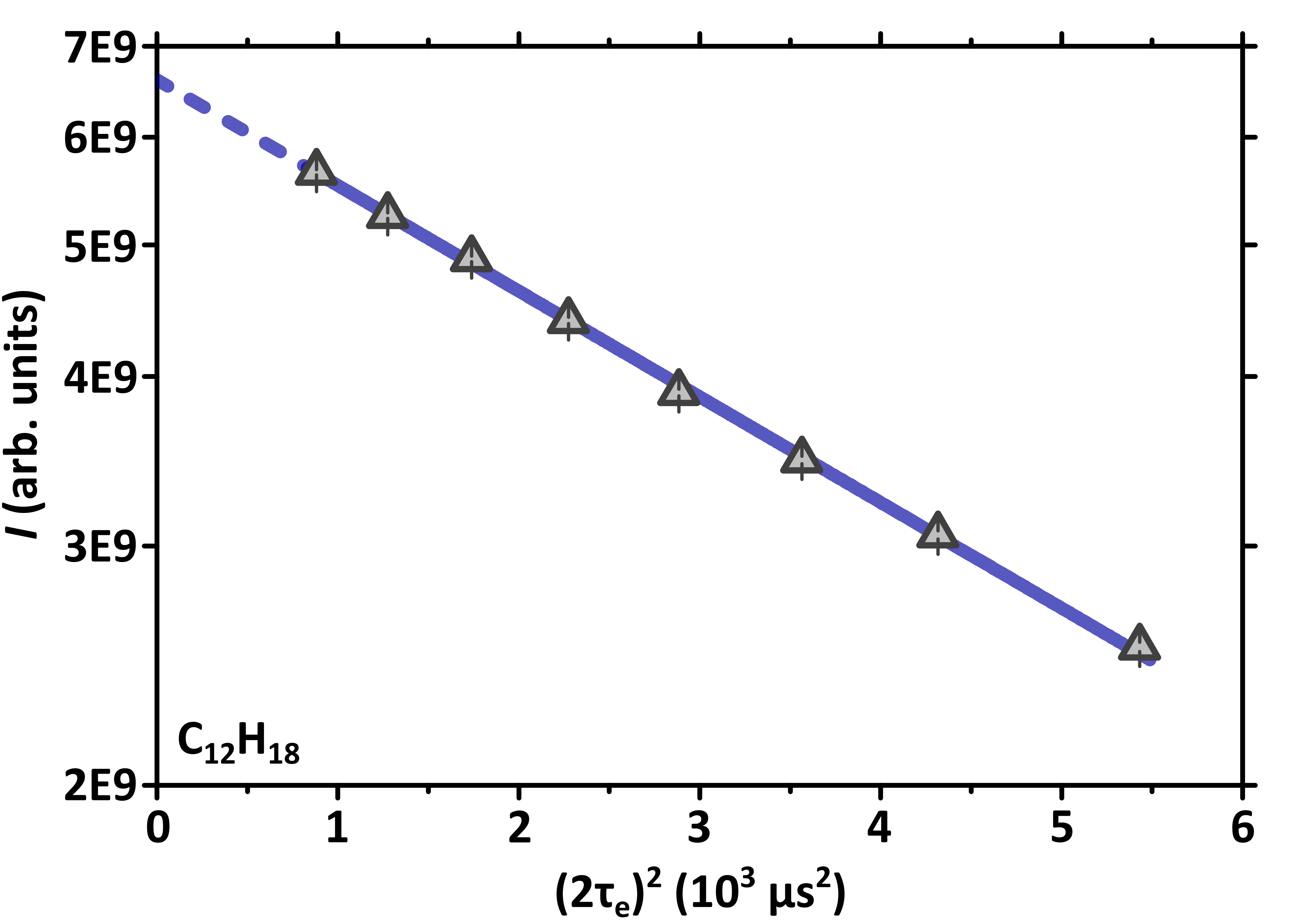}
	\caption{\label{FigDecay} Left-hand panel: representative spin-echo amplitude decay as a function of twice the time separation between the pulses for the TiSe$_{2}$H$\ped{x}$ crystals ($\tau\ped{pulse} = 1.1 \; \upmu$s, $360$ repetitions). Right-hand panel: representative spin-echo amplitude decay as a function of the square of twice the time separation between the pulses for the C$_{12}$H$_{18}$ reference sample ($\tau\ped{pulse} = 1.7 \; \upmu$s, $160$ repetitions). The continuous lines are best-fitting exponential-decay (TiSe$_{2}$H$\ped{x}$) and Gaussian-decay (C$_{12}$H$_{18}$) curves, while the dashed line highlight the extrapolation back to $2\tau\ped{e} = 0$ based on the fitting curves.}
\end{figure}
The acquisition window is adjusted so that only the second half of the spin-echo signal is detected (see Supplementary Fig.\,\ref{FigNMRdiag}). The resulting signal undergoes fast Fourier transform and the transformed signal is numerically integrated, resulting in the value $I$. The whole process is repeated for different $\tau\ped{e}$ values and, eventually, the value of $I$ is plotted as a function of $2\tau\ped{e}$. Representative curves for both TiSe$_{2}$H$\ped{x}$ and C$_{12}$H$_{18}$ are reported in the left-hand and right-hand panels of Supplementary Fig.\,\ref{FigDecay}, respectively. The experimental data are best-fitted based on exponential-decay (TiSe$_{2}$H$\ped{x}$) and Gaussian-decay (C$_{12}$H$_{18}$) curves, making it possible to extrapolate the integrated signal amplitude back to the limit $2\tau\ped{e} = 0$. This extrapolated value is independent on any signal suppression due to sample-specific $T_{2}$ processes and is directly related to the number of resonating {}$^{1}$H nuclear magnetic moments -- as such, this is the relevant quantity for the proton quantification procedure. Incidentally, we stress that both the TiSe$_{2}$H$\ped{x}$ crystals and the reference C$_{12}$H$_{18}$ sample were wrapped in teflon tape and we ensured that both the empty coil and the teflon tape did contribute with negligible signal amplitude. We repeated the procedure described above for different RF pulse durations, aiming at the maximization of the extrapolated $I$ value at $2\tau\ped{e} = 0$ (achieved for a $\pi/2$ pulse, with duration $\tau\ped{\pi/2}$).

\begin{figure}[h] 
	\centering
	\includegraphics[width=0.48\textwidth]{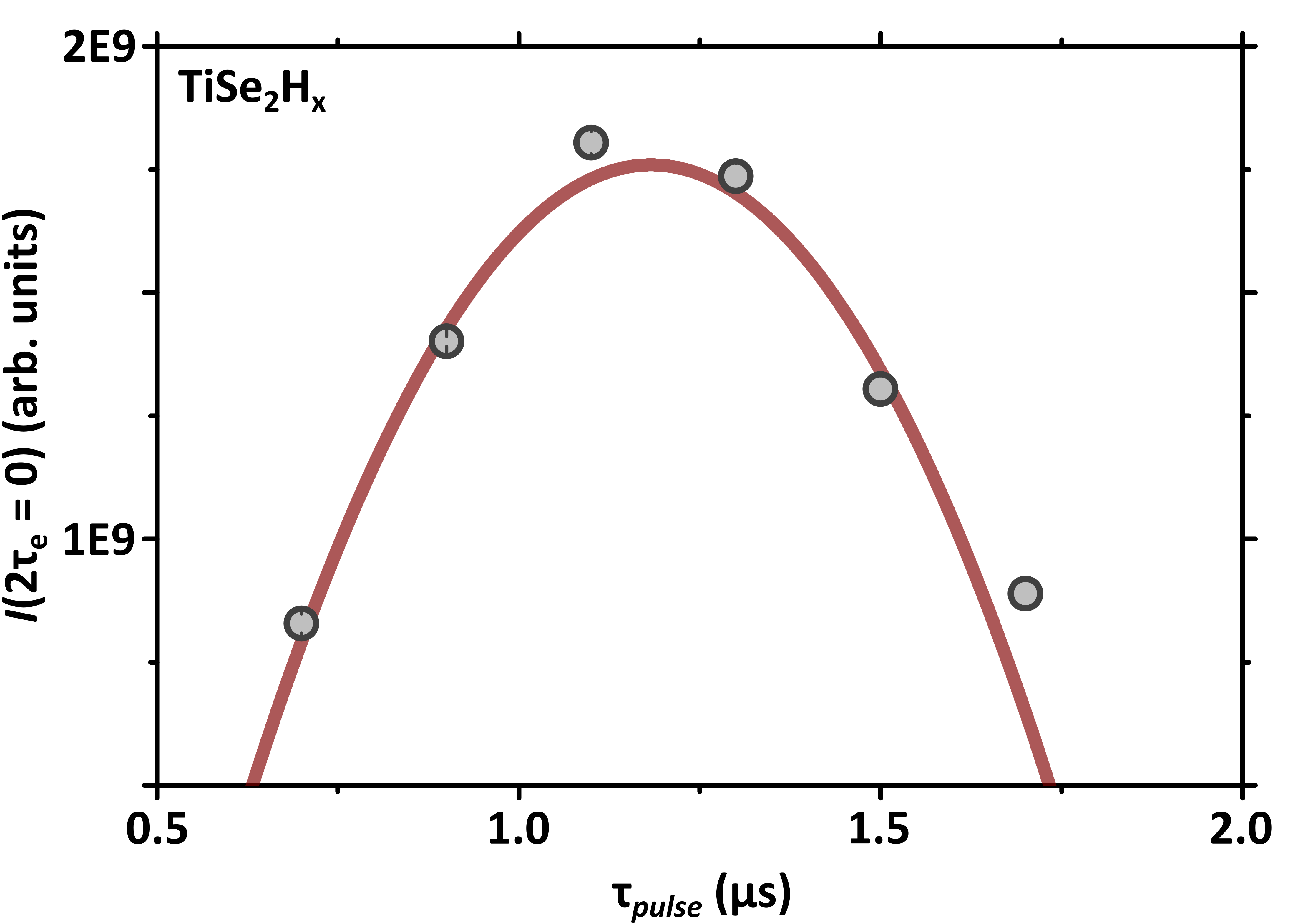} \includegraphics[width=0.48\textwidth]{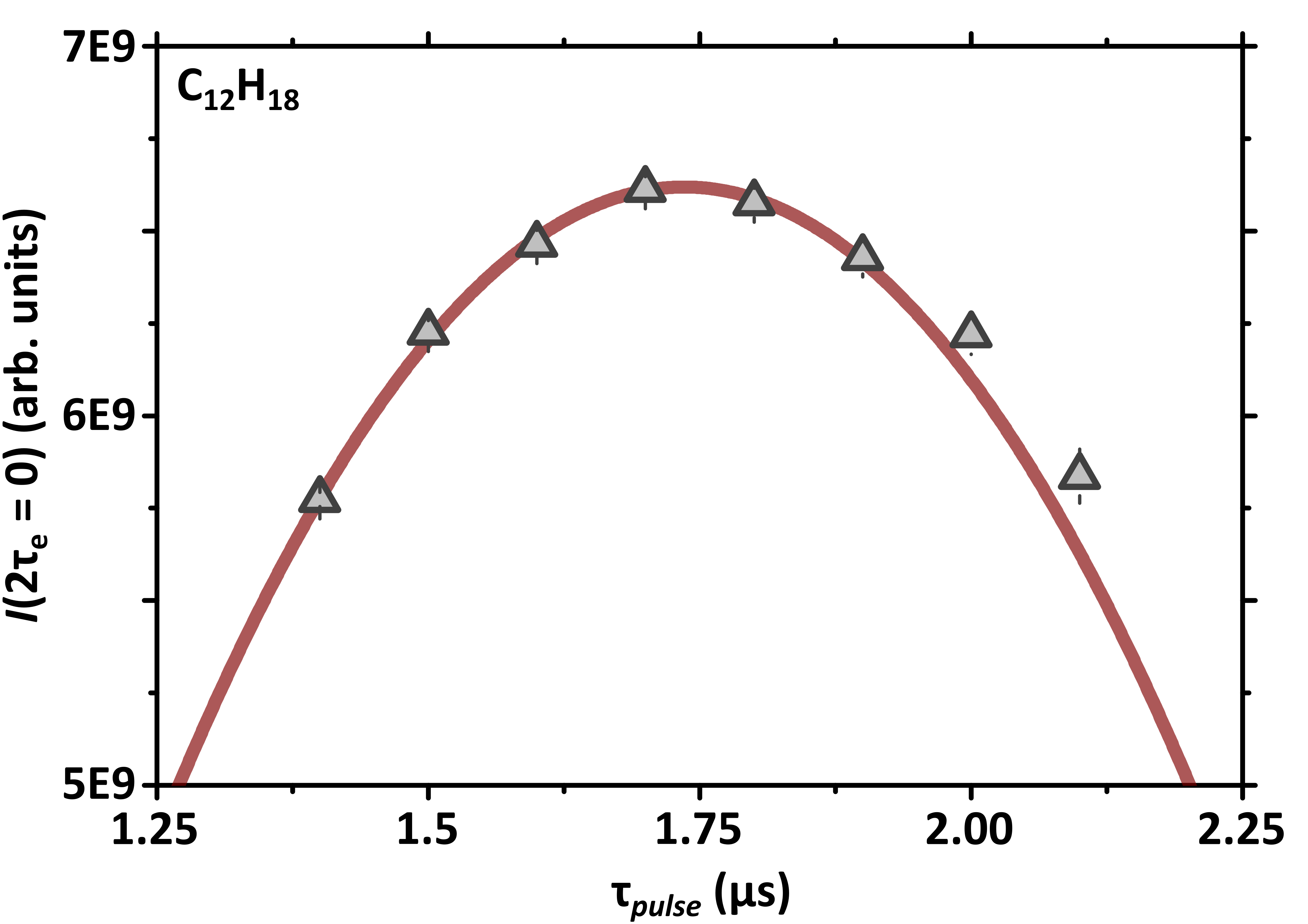}
	\caption{\label{FigMaxInt} Extrapolated $I$ value at $2\tau\ped{e} = 0$ (see Supplementary Fig.\,\ref{FigDecay}) as a function of the pulse length. The continuous lines are best-fitting parabolic curves.}
\end{figure}
The results for TiSe$_{2}$H$\ped{x}$ and C$_{12}$H$_{18}$ are reported in the left-hand and right-hand panels of Supplementary Fig.\,\ref{FigMaxInt}, respectively. The ratio between the maximum $I$ values for TiSe$_{2}$H$\ped{x}$ and C$_{12}$H$_{18}$ should be equal to the ratio between the number $N\ped{p}$ of protons in the two samples. However, we notice that the maxima for the two samples correspond to two different calibrated pulse durations ($\tau\ped{\pi/2} = 1.18 \pm 0.05 \; \upmu$s for TiSe$_{2}$H$\ped{x}$ and $\tau\ped{\pi/2} = 1.74 \pm 0.04 \; \upmu$s for C$_{12}$H$_{18}$), suggesting different quality factors for the circuit when the different samples are inserted in the coil. In particular, a higher quality factor is associated to the circuit transferring power more effectively, i.e., with a shorter $\tau\ped{\pi/2}$. At the same time, based on the reciprocity principle, a higher quality factor also corresponds to a higher coil sensitivity for the signal detection. Overall, neglecting the small discrepancies in the geometrical filling factor in the two measurements, we have that\,\cite{Hou76,Mo09}
\begin{equation}\label{EqProtonNumberRatio}
	\frac{\left(N\ped{p}\right)_{\rm{TiSe}_{2}\rm{H}\ped{x}}}{\left(N\ped{p}\right)_{\rm{C}_{12}\rm{H}_{18}}} = \frac{\left(\tilde{I}\ped{max} \times \tau\ped{\pi/2}\right)_{\rm{TiSe}_{2}\rm{H}\ped{x}}}{\left(\tilde{I}\ped{max} \times \tau\ped{\pi/2}\right)_{\rm{C}_{12}\rm{H}_{18}}}.
\end{equation}
In the equation above, we have defined $\tilde{I}\ped{max}$ as the maximum value $I\ped{max}$ (see Supplementary Fig.\,\ref{FigMaxInt}) normalized by the number of repetitions of the sequence. We can calculate $\left(N\ped{p}\right)_{\rm{C}_{12}\rm{H}_{18}}$ based on the sample mass ($3.16 \pm 0.05$\,mg) and the associated molar mass ($162.27$\,g/mol$_{\rm{C}_{12}\rm{H}_{18}}$) and, specifically, we have
\begin{equation}
	\left(N\ped{p}\right)_{\rm{C}_{12}\rm{H}_{18}} = 18 \times \left(\frac{(3.16 \pm 0.05) \times 10^{-3} \; \rm{g}}{162.27 \; \rm{g/mol_{\rm{C}_{12}\rm{H}_{18}}}} \times N\ped{A}\right) = (2.11 \pm 0.03) \times 10^{20}
\end{equation}
with $N_{A}$ the Avogadro constant. Now, $\left(N\ped{p}\right)_{\rm{TiSe}_{2}\rm{H}_{x}}$ -- i.e., the total number of resonating {}$^{1}$H nuclear magnetic moments inside the TiSe$_{2}$H$\ped{x}$ sample -- is the only unknown quantity in Supplementary Eq.\,\eqref{EqProtonNumberRatio}. Based on the experimental values extracted from Supplementary Fig.\,\ref{FigMaxInt}, we have a direct quantification of $\left(N\ped{p}\right)_{\rm{TiSe}_{2}\rm{H}_{x}}$ as
\begin{equation}
	\left(N\ped{p}\right)_{\rm{TiSe}_{2}\rm{H}_{x}} = (1.7 \pm 0.2) \times 10^{19}.
\end{equation}
This is more easily expressed as x, i.e., the number of {}$^{1}$H nuclear magnetic moments per crystallographic cell of TiSe$_{2}$H$\ped{x}$. The number of crystallographic cells in the TiSe$_{2}$H$\ped{x}$ crystals $\left(N\ped{c}\right)_{\rm{TiSe}_{2}\rm{H}_{x}}$ can be approximated based on the sample mass ($2.90 \pm 0.06$\,mg) and the molar mass of pristine TiSe$_{2}$ ($205.81$ g/mol$_{\rm{TiSe}_{2}}$) as
\begin{equation}
	\left(N\ped{c}\right)_{\rm{TiSe}_{2}\rm{H}_{x}} \simeq \frac{(2.90 \pm 0.06) \times 10^{-3} \; \rm{g}}{205.81 \; \rm{g/mol_{\rm{TiSe}_{2}}}} \times N\ped{A} = (8.5 \pm 0.2) \times 10^{18}.
\end{equation}
Overall, our final result is
\begin{equation}
	\mathrm{x} = \frac{\left(N\ped{p}\right)_{\rm{TiSe}_{2}\rm{H}_{x}}}{\left(N\ped{c}\right)_{\rm{TiSe}_{2}\rm{H}_{x}}} = 2.0 \pm 0.3.
\end{equation}

\bigskip
\bigskip

\section*{Supplementary Note 3: Raman spectra of hydrogen-doped titanium diselenide}

As mentioned in the Main Text, the inclusion of H dopants removes the symmetries of the systems making all modes Raman active. Although formally all modes are Raman active, it is not certain that their intensity is actually detectable.
Since after the inclusion of H atoms the systems are metallic  (at least the system with only one H), it is not possible to calculate the theoretical intensity of the peaks. It is however possible to analyse the phononic eigenvectors against the Raman-active ones in the pristine system searching for a possible relation in atomic displacements. 
{\color{blue}From this analysis we can directly justify the appearance of the peaks at $\sim$\,156 and $\sim$\,260\,cm$^{-1}$ in the experimental Raman spectra of H\ped{2}TiSe\ped{2} (Main Fig.\,1g). Specifically, the presence of the two peaks can be accounted in part by the configuration involving a single H atom in the "bridge position" modifying the Se vibrational properties, and in part by configuration where two H atoms are included in the TiSe\ped{2} structure, one in the VdW gap and the other one in the TiSe$_2$ trilayer (see also Supplementary Note 7).

Accounting for the broad band at $\sim$\,2900\,cm\apex{-1} on the other hand is more complex. In general, molecular hydrogen shows a high-frequency stretching mode which falls around $\sim$\,4000\,cm\apex{-1} and strongly depends on the structural environment in which H$_2$ is hosted\,\cite{FuteraJPCC2017}. Our calculations indicate that the H$_2$ vibrational frequency lies at $\sim$\,4100\,cm\apex{-1} when the molecule is isolated; but when the H$_2$ is intercalated in the TiSe\ped{2} matrix (i.e, the H\ped{2}TiSe\ped{2} phase where the H\ped{2} molecule lies in the VdW gap with the H--H bond parallel to the TiSe$_2$ plane) its vibrational frequency softens down to $\sim$\,3700\,cm\apex{-1}, due to the structural and metallic electronic environment.
This result is in line with what observed for other systems containing molecular hydrogen, in which the interstitial H$_2$ molecule shows a strongly renormalized frequency down to $\sim$\,3000\,cm\apex{-1}\,\cite{FuteraJPCC2017, OkamotoPRB1997}.
Based on this result, the very broad band roughly centered at $\sim$2900\,cm\apex{-1} in the experimental spectrum can be naturally interpreted as originating from H$_2$ molecules confined in the TiSe$_2$ matrix, with renormalized phonon frequency. 
Considering that our estimation of the frequency shift was obtained for H$_2$ in the most stable site found in TiSe$_2$, disorder effects related to multiple-site intercalation, interaction with other molecules due to inhomogeneities, temperature effects and/or interactions with free atomic hydrogens will result in a broad feature statistically including different frequency shifts (depending on structural environment and electronic doping).
Overall, the experimental observation of the additional low-frequency peaks and of the broad band at high frequencies in the Raman spectrum is therefore a further indication of the presence of disorder in the hydrogen concentration, phase and distribution in the intercalated TiSe$_2$ samples.}

\clearpage

\section*{\label{ssec:Hall}Supplementary Note 4: Temperature-dependent transverse resistivity measurements}

\begin{figure}[h]
	\includegraphics[width=\textwidth]{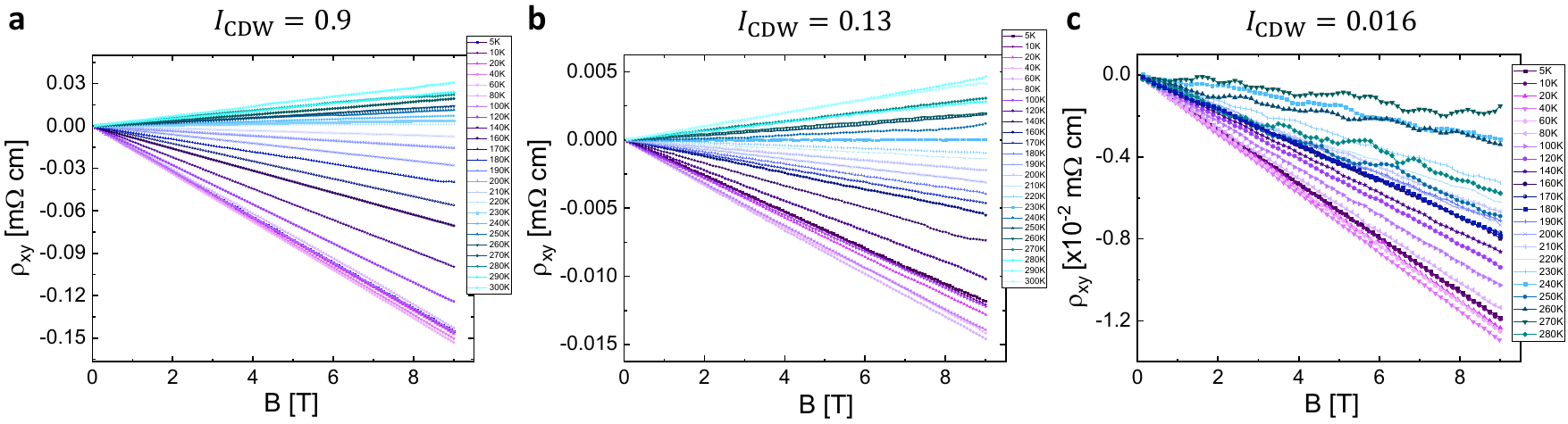}
	\caption{\label{fig:Hall}
		Antisymmetrized transverse resistivity $\rho\ped{xy}$ as a function of the out-of-plane magnetic field $B$ measured at different temperatures $T$ between $5$ and $300$\,K in three TiSe\ped{2} samples with different charge-density wave intensities $I\ped{CDW}$ (see Main Text for details). \textbf{a}, a pristine TiSe\ped{2} sample with $I\ped{CDW} = 0.9$. \textbf{b}, an intermediately-doped H\ped{x}TiSe\ped{2} sample with $I\ped{CDW} = 0.13$. \textbf{c}, a fully-doped H\ped{2}TiSe\ped{2} sample with $I\ped{CDW} = 0.016$. The resulting $T$-dependence of the Hall coefficient is plotted in Fig.\,2e of the Main Text.
	}
\end{figure}

Magnetic field ($B = \mu_0H$)-dependent transverse resistivity ($\rho\ped{xy}$) measurements were carried out at temperatures ranging between 5\,K and 300\,K on three representative TiSe\ped{2} crystals as discussed in the Main Text and in the Methods. Supplementary Fig.\,\ref{fig:Hall} reports the full sets of temperature-dependent antisymmetrized $B-\rho\ped{xy}$ curves. The corresponding values of $R\ped{H}$ reported in the Main Text are obtained by linearly fitting the experimental data at a given temperature in the entire $B$ range.
{\color{blue}Additionally, all $B-\rho\ped{xy}$ curves exhibit good linear behaviour up to 9\,T. This indicates that, if H\ped{x}TiSe\ped{2} is a multi-band superconductor as suggested by the observed upturn in the temperature-dependence of the critical magnetic field, its electronic structure does not support the ideal conditions for the observation of a non-linear Hall effect, namely a first band with high carrier density and low mobility and a second band with low density and high mobility\,\cite{DingNL2022}. This is  consistent with existing data in undoped TiSe\ped{2}\,\cite{KnowlesPRL2020}. However, the limited range of magnetic fields accessible in our cryostat do not allow us to discern whether the possible multi-band superconductivity would be supported by two bands of comparable mobility and different density, or two bands of comparable density and different mobility, or a first band with high density and high mobility and a second band with low density and low mobility.}

\section*{\label{ssec:SC_vs_B}Supplementary Note 5: Additional temperature-dependent magnetotransport measurements in the superconducting phase of \texorpdfstring{H\ped{2}TiSe\ped{2}}{H2TiSe2}}

\begin{figure}[h]
\includegraphics[width=0.7\textwidth]{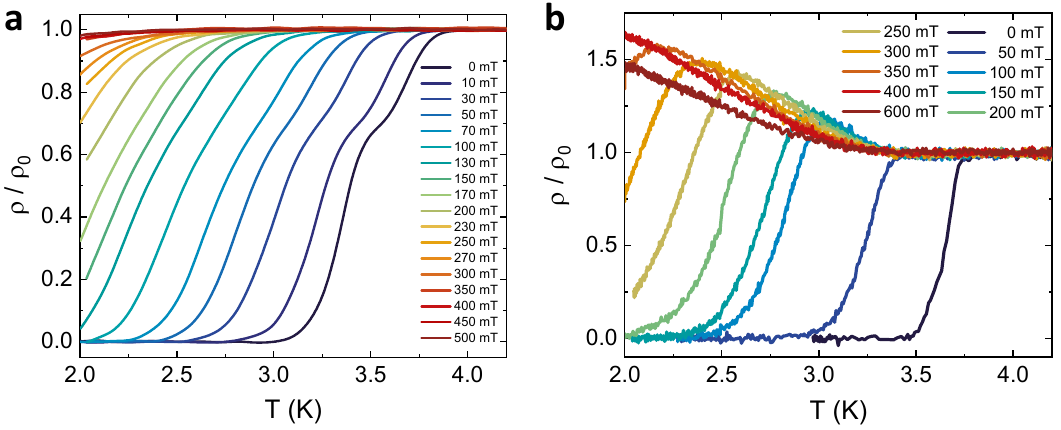}
\caption{\label{fig:SCvsB}
	Resistivity $\rho$ of two fully-doped H\ped{2}TiSe\ped{2} crystals as a function of temperature $T$ in the superconducting transition, normalized by its normal-state value $\rho_0$, for increasing magnetic field $H$ applied orthogonal to the crystal $ab$ plane.
}
\end{figure}

Supplementary Fig.\,\ref{fig:SCvsB} shows the temperature-dependent resistivity of two additional H\ped{2}TiSe\ped{2} crystals across the superconducting transition, at different values of the magnetic field applied perpendicular to the $ab$ planes. The resistivity is normalized to its value in the normal state immediately above the SC transition.

\clearpage

\section*{\label{ssec:samp_prep}Supplementary Note 6: Sample preparation for muon spectroscopy measurements}
{\color{blue}
A set of H$_{2}$TiSe$_{2}$ single crystals belonging to the same batch was glued using cryogenic heat-sinking varnish (GE 7031) on a thin (25-$\upmu$m) Cu foil, so as to realize a mosaic (surface exposed to the muon beam $\sim 20$ mm$^2$ -- see Supplementary Fig.\,\ref{fig:usrsamples}). The resulting sample consisted 
of crystals with aligned $c$ axes, but with random orientation of the in-plane $a$ and $b$ axes. A few layers of crystals were necessary in order to achieve a minimum thickness of about 0.2\,mm. Such average sample thickness and the presence of a 100-$\upmu$m-thick high-purity silver degrader on the sample surface ensure that muons are implanted in the sample (and do not pass through it). 
\begin{figure}[h]
\includegraphics[width=0.7\textwidth]{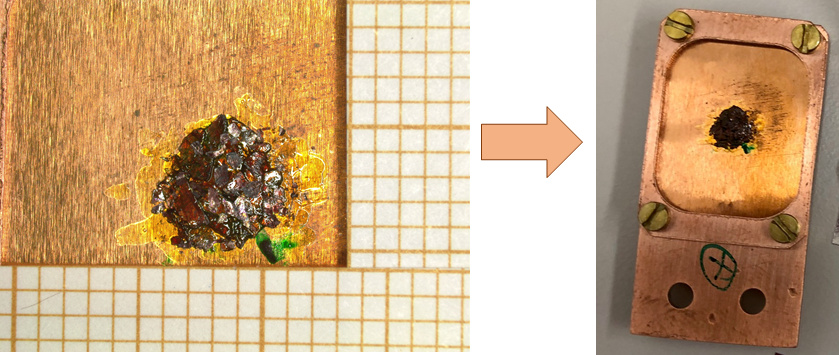}
\caption{\label{fig:usrsamples}
{\color{blue}Left: mosaic of H$_{2}$TiSe$_{2}$ single crystals glued on a copper foil (the 1-mm graph paper is juxtaposed for a size comparison). Right: the sample glued on the copper foil is mounted in the copper sample holder of the Dolly spectrometer.}}
\end{figure} 
}
\section*{\label{ssec:zfmu}Supplementary Note 7: Zero-field muon spin rotation measurements}
\begin{figure}[tbh]
\includegraphics[width=0.35\textwidth]{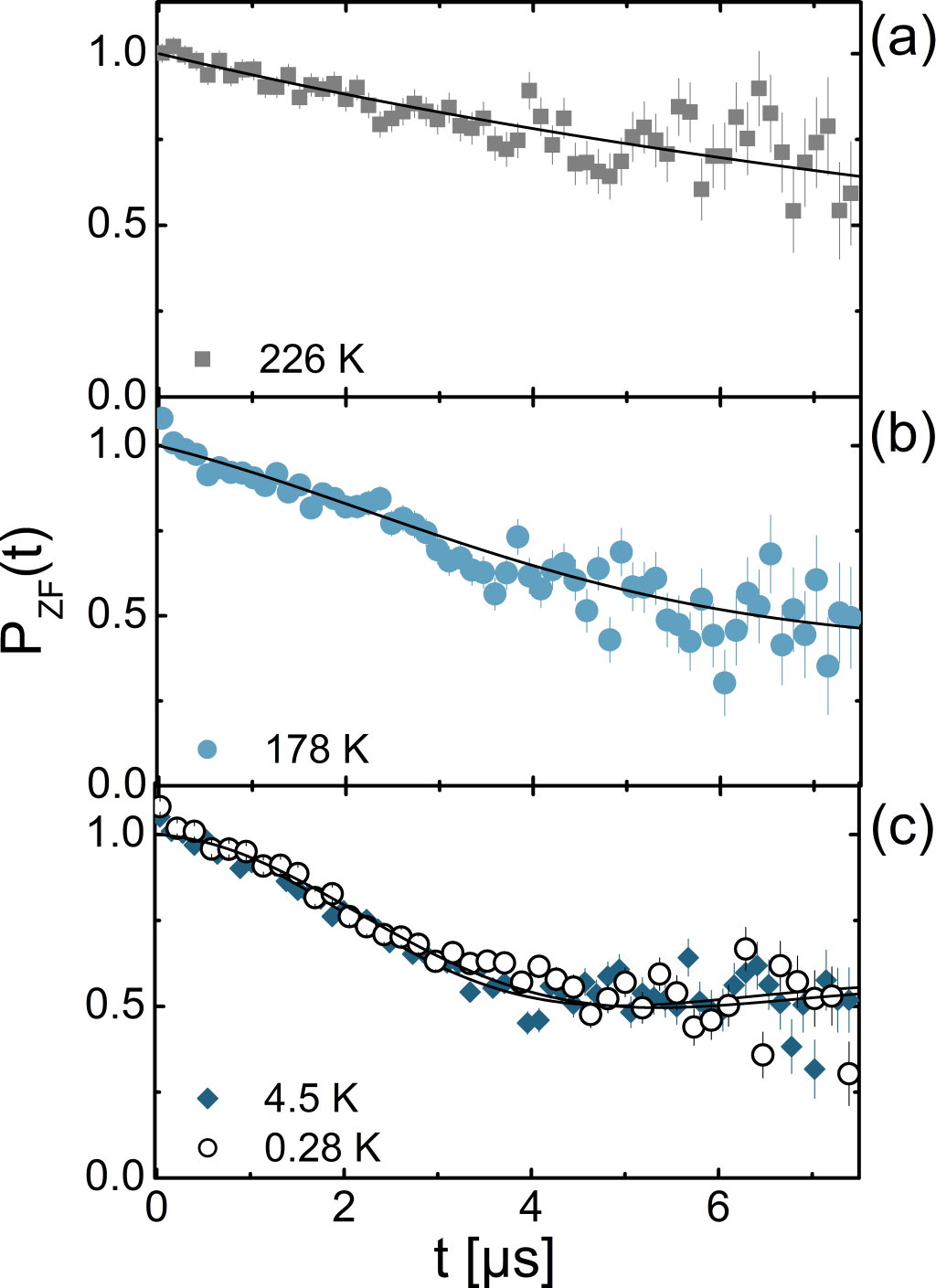}
\caption{\label{fig:ZFpol}Time-domain zero-field muon-spin polarization at selected temperatures. The continuous lines represent fits to Supplementary Eq.~\eqref{eq:pt_zf}. See text for details.}
\end{figure}

Supplementary Fig.\,\ref{fig:ZFpol} shows a selected set of zero-field muon-spin relaxation (\zfmu) measurements whose aim was twofold: to detect a possible time-reversal symmetry breaking (TRSB), usually associated with unconventional pairing, and to evaluate the amount of hydrogen atoms effectively intercalated in the layered TiSe$_{2}$ structure. As to the former, by measuring the time-dependent polarization in a strictly zero field (the background magnetic fields being actively compensated) one can detect the tiny magnetic fields associated with the electrons paired in a spin-triplet state. The TRSB has been observed in Sr$_2$RuO$_4$~\cite{Luke1998} and related compounds\,\cite{Shiroka2012}, as well as in some topological superconductors, such as Sr$_{0.1}$Bi$_2$Se$_3$\,\cite{Biswas2019}. In Supplementary Fig.\,\ref{fig:ZFpol}c, we show the time-dependent \zfmu\ polarization above and below the superconducting transition. Clearly, the two da\-ta\-sets fully overlap, thus excluding a possible increase of depolarization rate in the superconducting state due to spontaneous magnetic fields  
and ruling out any TRSB.

All the \zfmu\ datasets were fitted by means of the model below:
\begin{equation}\label{eq:pt_zf}
  P_\mathrm{ZF}(t) =  f_\mathrm{tail} \cdot \exp(-\Lambda t) + \left(1-f_\mathrm{tail}\right) \cdot  G_\mathrm{KT}(t) \cdot \exp(-\nu t).
\end{equation}
Here, the long-time relaxing tail is due to the muons implanted in the silver degrader. Both $f_\mathrm{tail}$ and $\Lambda$ were fixed to the values determined by \tfmu\ measurements. 
Under some broad assumptions, the second term can be referred to the sample: $G_\mathrm{KT}(t)$ is the static Gaussian Kubo-Toyabe function reflecting the contribution of randomly-oriented nuclear magnetic moments. 
Such term is multiplied by a simple exponential decay which accounts for 
possible dynamic processes, such as muon- and/or hydrogen hopping (expected to occur at high temperatures~\cite{Hayano1979,Yaouanc2011}), with $\nu$ being the hopping rate.

\begin{figure}[tbh]
\includegraphics[width=0.35\textwidth]{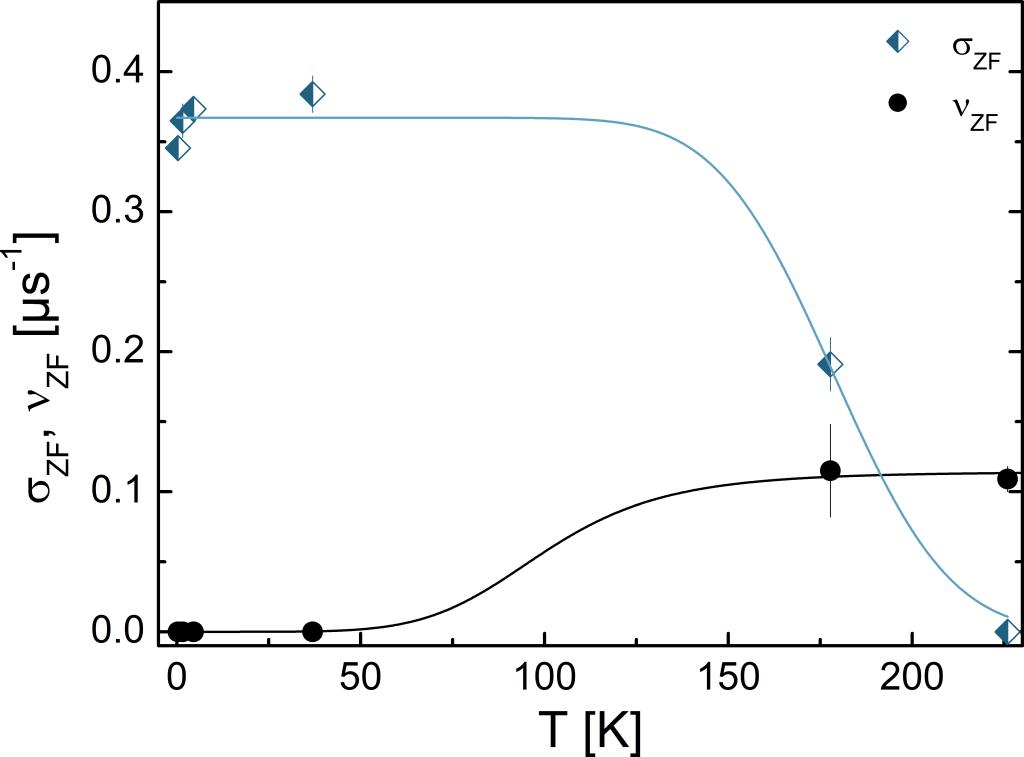}
\caption{\label{fig:ZF-Relax}Temperature dependence of the zero-field muon-spin depolarization rate $\sigma_\mathrm{ZF}$ and hopping rate $\nu_\mathrm{ZF}$ in H-doped TiSe$_2$. The lines are guides for the eyes.}
\end{figure}

Supplementary Fig.\,\ref{fig:ZF-Relax} presents the temperature evolution of the \zfmu\ depolarization and the hopping rates, as resulting from fitting the $P_\mathrm{ZF}(t)$ data to Supplementary Eq.~\eqref{eq:pt_zf}. First, we observe that our results agree closely (at least qualitatively) with those of ZrV$_{2}$H$_{x}$, a Laves-phase material~\cite{Soetratmo97}, where one can distinguish two 
different regimes. At low temperatures (below 50\,K), muons are static and show a relatively large depolarization rate, $\sigma_\mathrm{ZF} = 0.346(6)$\,$\upmu$s$^{-1}$. For $T > 50$\,K, $\sigma_\mathrm{ZF}$ decreases progressively because motional narrowing sets in. 
The latter is fully effective for $T > 180$\,K, where both a sharp decrease of $\sigma_\mathrm{ZF}$ and a corresponding increase of $\nu$ take place. As observed already in ZrV$_{2}$H$_{x}$, also in our case, muon motions seem to be 
highly correlated with those of the intercalated hydrogen atoms: most of the interstitial sites available  for muon hopping being occupied by the hydrogen. When, at high temperatures, hydrogen too starts to  diffuse, 
``new'' interstitial sites become available for the muon diffusion, thus explaining the sharp decrease in depolarization rate observed for $T > 180$\,K. Experimental data on ZrV$_{2}$H$_{x}$ suggest also that, in such system, muon hopping is 
slower than hydrogen hopping~\cite{Hempelmann89,Soetratmo97}, a counter-intuitive result considering the ninefold lower mass of muon compared to that of proton. 
To date, there are no clear explanations for this apparently contradictory result. 

Interestingly, it was also shown that, at low temperatures, the $\sigma_\mathrm{ZF}$ value increases with increasing H content~\cite{Hempelmann89}, reaching 0.37\,$\upmu$s$^{-1}$ for $x = 4.8$. Since this decay rate agrees with that observed in our case, it suggests a high degree of H intercalation also in TiSe$_{2}$, thereby confirming -- at least qualitatively -- the results of proton-content quantification by ${}^{1}$H-NMR measurements discussed above.

\clearpage

\section*{\label{ssec:DFT}Supplementary Note 8: Additional details on the density functional theory calculations}

\subsection{Electronic properties as a function of increasing hydrogen doping}

\begin{figure}[tbh]
\includegraphics[width=0.8\textwidth]{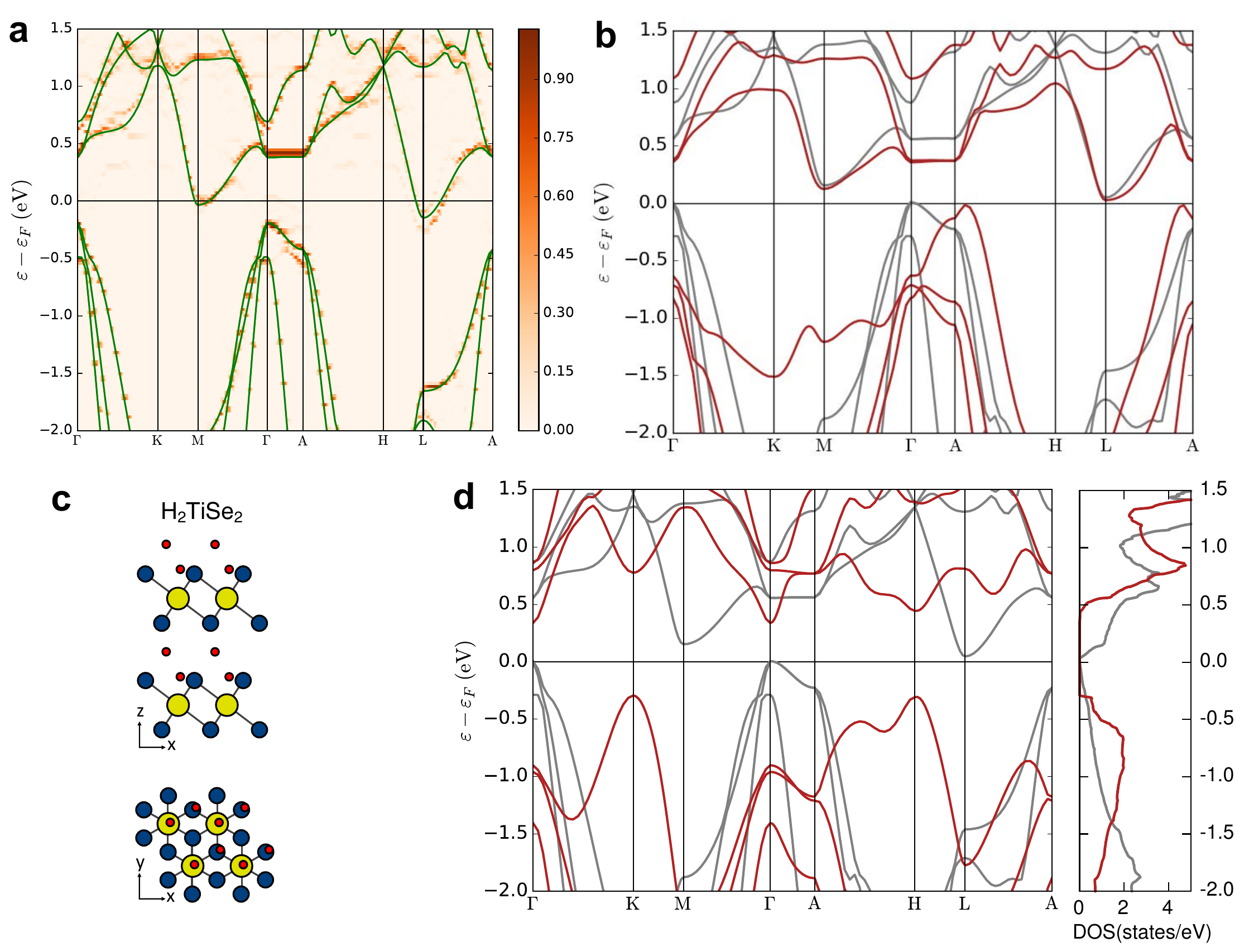}
\caption{\label{fig:H2TiSe2bands}
\textbf{a}, Unfolded electronic band structure of $1T$-H\ped{0.125}TiSe\ped{2} shown as orange scale, the intensity following the spectral weight of the band, compared with the 0.1$e^-$ doped $1T$-TiSe\ped{2} dispersion obtained in the jellium approximation (green line).
\textbf{b}, Electronic band structure of 1$T$-H$_2$TiSe$_2$ where the H dopants are intercalated as H\ped{2} molecules (red) compared with the pristine 1$T$-TiSe$_2$ dispersion (gray).
\textbf{c}, Ball-and-stick model of the alternative H$_2$TiSe$_2$ adsorption configuration, metastable with respect to the H\ped{2} molecule formation inside the van der Waals gap.
\textbf{d}, Electronic band structure and density of states of $1T$-H\ped{2}TiSe\ped{2} in the configuration reported in \textbf{c} (red) compared with the pristine $1T$-TiSe\ped{2} dispersion (gray).
}
\end{figure}

The random-search algorithm analysis highlighted the existence of several different metastable intercalation sites, indicating that hydrogen can easily be ``trapped" (at $T=0$) in different local minima. The lowest-energy phase discovered features the H-atom located in the van der Waals gap, at the ``bridge" position between two Se-atoms belonging to different planes. Interestingly, a metastable phase where the H atom is located inside the TiSe$_2$ trilayer is found very close in energy to the lowest-energy one ($\Delta E\sim0.06$~eV).
The band structure of 1$T$-H$_{0.125}$TiSe$_2$, unfolded in the ($1\times1\times1$) Brillouin Zone (Supplementary Fig.\,\ref{fig:H2TiSe2bands}a) strongly resembles that of ideal TiSe$_2$, but with the Fermi level now shifted in the conduction band populating the electron valley at the $L$-point of the Brillouin zone, with an estimated electron doping is of about 0.1$e^-$ with respect to the pristine 1$T$-TiSe$_2$. Thus at low-doping regime, H acts as an electron-donor, without appreciable changes of the band structure.

Following the experimental indications for large H doping, the hydrogen concentration was then increased to $\mathrm{x}=1$ and $\mathrm{x}=2$. At $\mathrm{x}=1$, the obtained intercalation sites are in line with the dilute case, but the band structure becomes strongly modified (as shown in Fig.\,5b of the Main Text).
At this higher doping level hydrogen still acts as an electron donor, but it now induces a strong band reconstruction lowering the Ti-$d_{z^2}$ which becomes half-filled.
A nominal doping of $\mathrm{x}=2$ (i.e. considering two inequivalent H atoms per unit cell, in line with the H content determined in fully-doped TiSe\ped{2} crystals via NMR) results in the formation of the H$_2$ molecule (laying in the VdW gap with the H--H bond parallel to the TiSe$_2$ plane) and with a negligible effect on the band structure of the pristine TiSe$_2$ (see Supplementary Fig.\,\ref{fig:H2TiSe2bands}b). 
However, the random search provides other (metastable) phases where H atoms remain in the atomic form and are bonded to the Se or Ti atoms, thus avoiding the formation of the molecule.
The electronic structure of one of these metastable phases, in which one H atom is intercalated in the vdW gap and the other within the trilayer (Supplementary Fig.\,\ref{fig:H2TiSe2bands}c), shows a transition to an insulating phase as the Ti-$d_{z^2}$ becomes completely filled (Supplementary Fig.\,\ref{fig:H2TiSe2bands}d).

\subsection{Dynamical stabilization of the metallic 
\texorpdfstring{H$_1$TiSe$_2$}{H1TiSe2} phase via the application of external pressure}

As discussed in the Main Text, all the single-phase structural models discussed above necessarily turn out to be dynamically unstable, as verified by the calculation of their phonon dispersion relations (not shown). This instability, which is likely resolved in the actual samples thanks either to the formation of H superstructures coupled with TiSe$_2$ CDW distortions, or to the strong H disorder, makes a calculation of their dynamical properties from first principles computationally unfeasible.

\begin{figure}[tbh]
\includegraphics[width=0.75\textwidth]{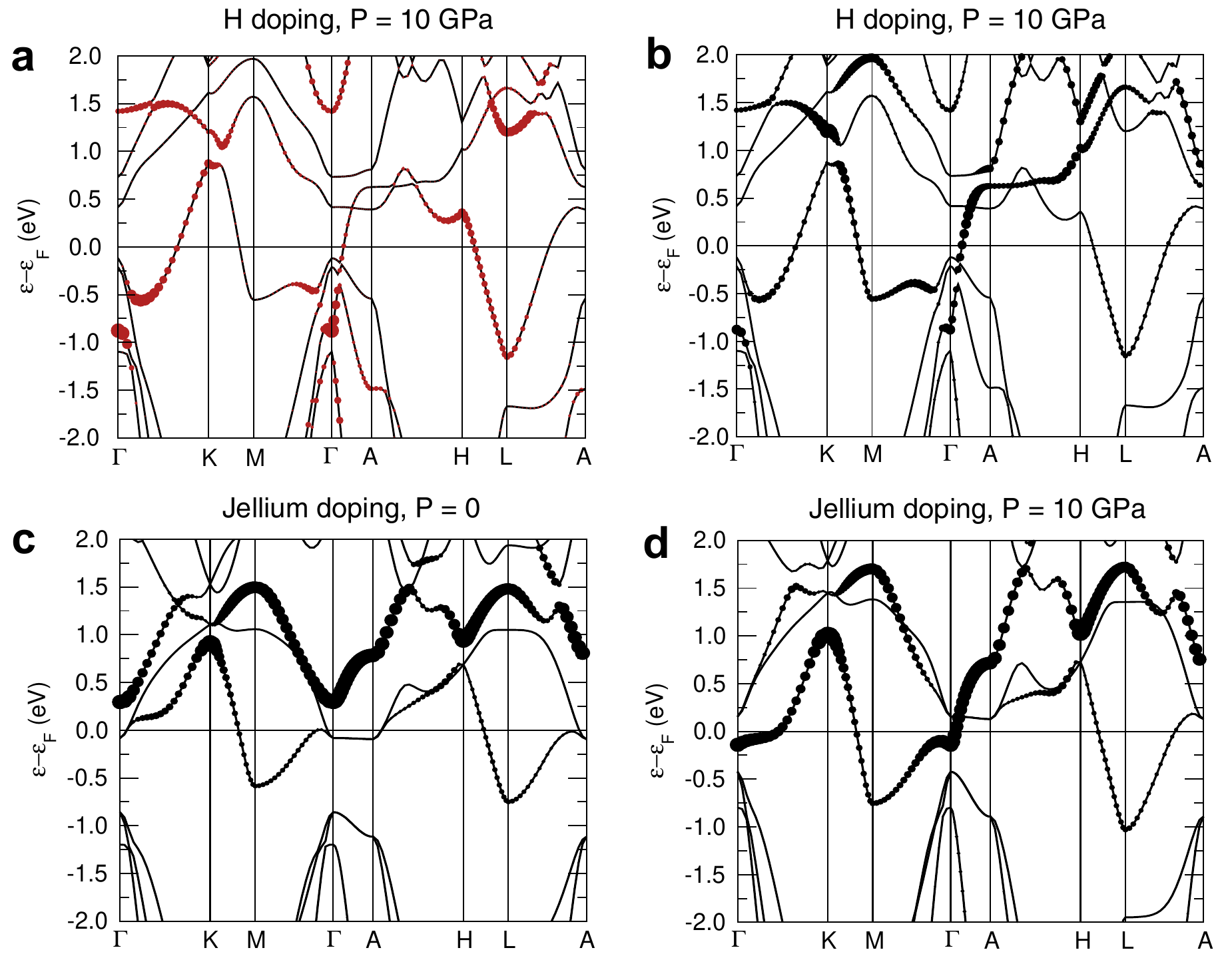}
\caption{\label{fig:pressurization}
\textbf{a}-\textbf{b}, Electronic band structure of $1T$-H$_1$TiSe$_2$ under the application of a pressure $\simeq10$~GPa.
In \textbf{a}, the size of the red circles is proportional to the H character of the eigenvalues.
In \textbf{b}, the size of the black circles is proportional to the $d_{z^2}$ character of the eigenvalues.
\textbf{c}, same as \textbf{b} for jellium-doped $1T$-TiSe$_2$ at 0~GPa.
\textbf{d}, same as \textbf{b} for jellium-doped $1T$-TiSe$_2$ at 10\,GPa.
}
\end{figure}

Nevertheless, we conceived a computational experiment which allows us to have solid first-principles predictions of the superconducting phase even in the high density regime. It consists in artificially stabilizing the metallic $1T$-H$_1$TiSe$_2$ phase (in which the Ti-$d_{z^2}$ band is partially filled) by slightly compressing the unit cell, so as to stiffen the short-range force constants and lift the phononic instabilities, without significantly changing the electronic properties of the material. This approach, already exploited to remove the CDW distortion in pure $1T$-TiSe$_2$\,\cite{PhysRevLett.106.196406} and reduce anharmonic effects in H-doped palladium alloys\,\cite{VocaturoJAP2022}, can here be used to obtain a dynamically stable high-doping phase.

Supplementary Fig.\,\ref{fig:pressurization}a-b show the electronic band structure of the $1T$-H$_1$TiSe$_2$ phase (solid black lines) once the lattice constants are reduced by 4\%  (that corresponds to an external pressure of $\sim 10$\,GPa). Despite a slight reduction in the bandwidths, the pressurized band structure closely resembles that computed at 0\,GPa and shown in Fig.\,5b of the Main Text, confirming that at this applied external pressure, the main features of the electronic properties of $1T$-H$_1$TiSe$_2$ are not strongly affected. Conversely, this same pressure is sufficient to make the system dynamically stable and bring it far from the critical region of the instability, as demonstrated by the phonon dispersion shown in Fig.5\,d of the Main Text. 
This confirms that this phase retains the important topology of the band structure of the doped phase, within a dynamically-stable phase. In particular, the electronic band structure shows a significant hybridization (depicted as red circles in Supplementary Fig.\,\ref{fig:pressurization}a) between the TiSe$_2$ bands and the H-derived orbitals. The largest percentage of the H component ($\sim7$\%) is found at the $\Gamma$ point and at an energy of $-0.6$\,eV, but is still sizeable ($\lesssim 5$\%) around the Fermi level.

\subsection{Comparison between the electronic structures of hydrogen-doped \texorpdfstring{TiSe$_2$}{TiSe2} and jellium-doped \texorpdfstring{TiSe$_2$}{TiSe2}}

The role played by the H dopants in determining the SC phase of $1T$-TiSe$_2$ can be further clarified by 
comparing the electronic properties of $1T$-H$_1$TiSe$_2$ with those of a $1T$-TiSe$_2$ where one electron per unit cell is added to the system together with a uniform compensating background (jellium model, see Methods), for both the experimental structure (0\,GPa) and the pressurized structure (10\,GPa). 
%
%
Supplementary Fig.\,\ref{fig:pressurization}c shows the electronic band structure at 0\,GPa of jellium-doped model, to be compared with that of H\ped{1}TiSe\ped{2} shown in Fig.\,5c of the Main Text. In both figures the size of the circles indicates the $d_{z^2}$ character of the bands. The comparison with the band structure of the TiSe$_2$ (grey lines in Fig.\,5b of the Main Text) immediately highlights the differences between the effects of actual H-insertion and pure H-driven charge doping. In particular, in the jellium model (Supplementary Fig.\,\ref{fig:pressurization}c) the doping is nearly rigid and the states at the Fermi level have a predominant in-plane character, while the $d_{z^2}$ band remains unoccupied. Instead, as discussed in the Main Text, the H insertion also leads to a significant deformation of the electronic dispersion and completely switches the orbital character of the bands which cross the Fermi level to $d_{z^2}$ states with a predominant out-of-plane character.

Upon the application of external pressure, the orbital character of the states at the Fermi level remains unaffected in the H-doped system (Supplementary Fig.\,\ref{fig:pressurization}b), thus indicating once again that the lattice compression necessary to stabilize the $1T$-H$_1$TiSe$_2$ does not give rise to any relevant change in the electronic and SC properties of the system. In the jellium-doped system, instead, pressure triggers a band inversion that leads, once again, to a partial filling of the highly-coupled $d_{z^2}$ band (Supplementary Fig.\,\ref{fig:pressurization}d) but, crucially, the system remains dynamically unstable. This indicates that the effect of a high concentration of H dopants in the TiSe$_2$ structure cannot simply be conceptualized as a combination of charge doping and chemical pressure.

\subsection{Dynamical and superconducting properties of jellium-doped \texorpdfstring{TiSe$_2$}{TiSe2}}
\begin{figure}[tbh]
\includegraphics[width=0.5\textwidth]{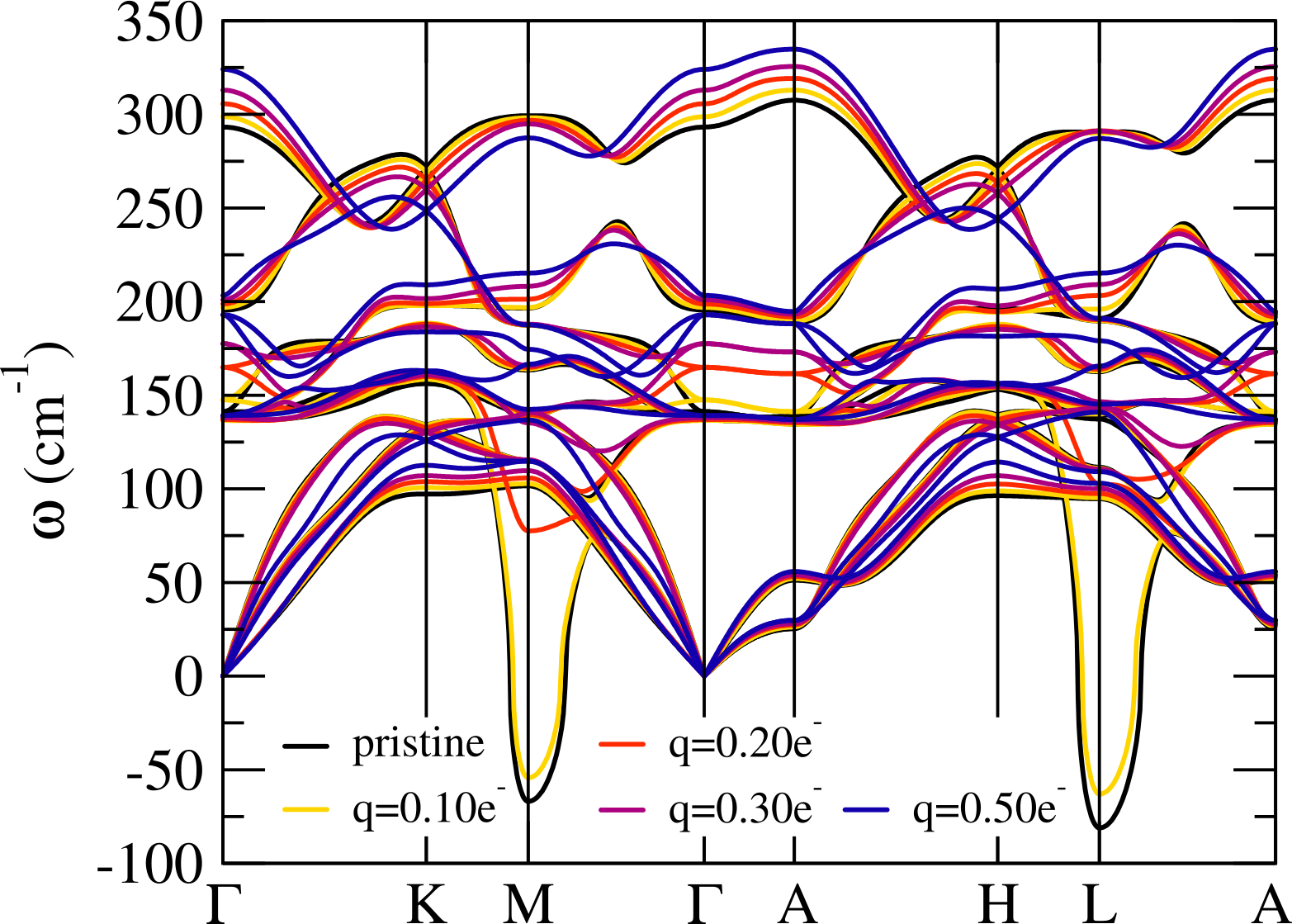}
\caption{\label{fig:TiSe2_ph_doping}
 Phonon dispersion relations of jellium-doped $1T$-TiSe$_2$ as a function of increasing electron doping.
}
\end{figure}

%
We verified that 1$T$-TiSe$_2$ doped in the jellium approximation can also support superconductivity. 
Supplementary Fig.\,\ref{fig:TiSe2_ph_doping} shows the phononic band structure of jellium-doped $1T$-TiSe\ped{2} as a function of electron doping, which evidences how,
%
%
as known for the 1$T$-TiSe$_2$ single layer\,\cite{PhysRevB.96.165404, Zhou2020}, electron doping is responsible for the dynamical stabilization of the system. In turn, the strong phonon softening associated with the disappearing phononic instability is responsible for the large electron-phonon coupling able to support a sizable $T\ped{c}$.
Specifically, when a charge doping of $\sim0.2e^-$ per unit cell is added to the $1T$-TiSe\ped{2} system, the first conduction band  
becomes populated and an electron-phonon coupling of $\lambda\sim 0.5$ is obtained, corresponding to a $T_c\sim 1.5$\,K. 
However, we stress again that -- unlike in the real H-doped system -- in this simplified ``toy" model high levels of electron doping ($\sim1e^-$ per unit cell) make the compressed TiSe$_2$ structure dynamically unstable. 
%
{\color{blue}Furthermore, the large H concentration levels estimated even in the least-doped samples (average $\mathrm{x}\sim0.4$ at the shortest gating time of $\sim 5$\,min) indicate that the jellium approximation does not correctly describe the superconducting properties of our H\ped{x}TiSe\ped{2} crystals at any gating time, and would be more suited to describe superconducting Cu- and Li-doped TiSe\ped{2}. 
This lack of applicability is particularly relevant in the likely case that, in samples at reduced H loading, the inhomogeneity in the H concentration leads to samples characterized by a percolating filamentary network of highly-doped regions embedded in a nearly undoped matrix.}

\clearpage
\bibliographystyle{naturemag}
\bibliography{bibliography}